\documentclass{iopart}
\usepackage{iopams}
\usepackage{setstack}
\usepackage{url}
\usepackage{bm}		
\usepackage{graphicx}	

\begin{document}

\title{Stereoscopic visualization in curved spacetime: seeing deep inside a black hole}

\author{Andrew~J~S~Hamilton$^{1,2}$ and Gavin~Polhemus$^{1,3}$}

\address{$^1$JILA, Box 440, U. Colorado, Boulder, CO80309, USA}
\address{$^2$Dept.\ Astrophysical \& Planetary Sciences, Box 391, U.~Colorado}
\address{$^3$Poudre High School, 201 Impala Drive, Fort Collins, CO 80521, USA}
\eads{\mailto{Andrew.Hamilton@colorado.edu}, \mailto{Gavin.Polhemus@colorado.edu}}

\newcommand{\simlt}{\alt}
\newcommand{\simgt}{\agt}
\newcommand{\Msun}{\rm{M}_\odot}

\newcommand{\dd}{{\rm d}}
\newcommand{\DD}{D}
\newcommand{\im}{\rmi}
\newcommand{\ee}{\rme}

\newcommand{\emit}{{\rm em}}
\newcommand{\obs}{{\rm obs}}
\newcommand{\ff}{{\rm ff}}

\newcommand{\bJ}{\bm{J}}
\newcommand{\bk}{\bm{k}}
\newcommand{\bL}{\bm{L}}
\newcommand{\bn}{\bm{n}}
\newcommand{\bu}{\bm{u}}

\newcommand{\unit}[1]{\, {\rm#1}}


\newcommand{\lensearthbinocularfig}{
    \begin{figure}
    \centering
    \includegraphics[width=4.5in]{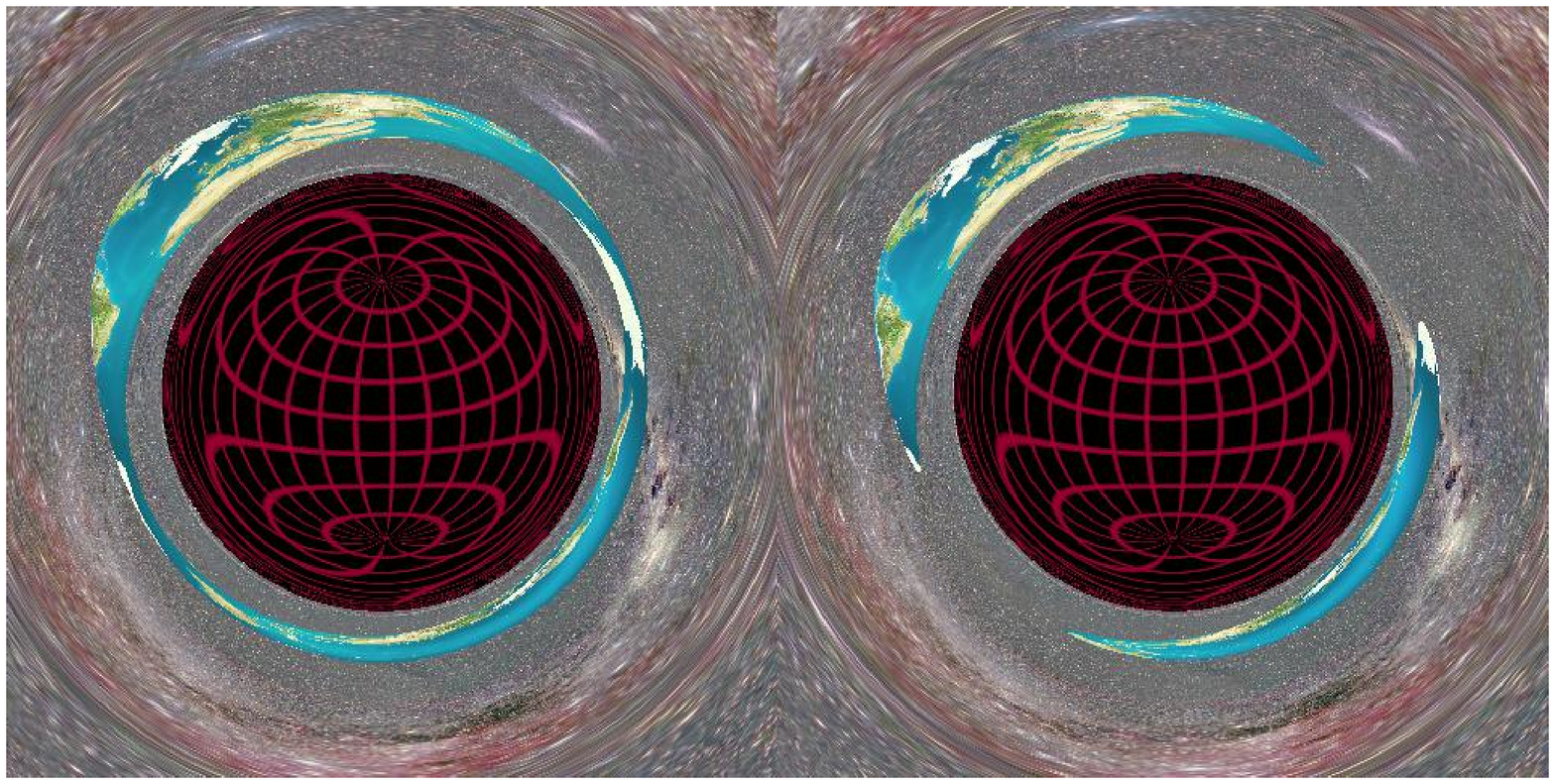}
    \vskip.5ex
    \includegraphics[width=4.5in]{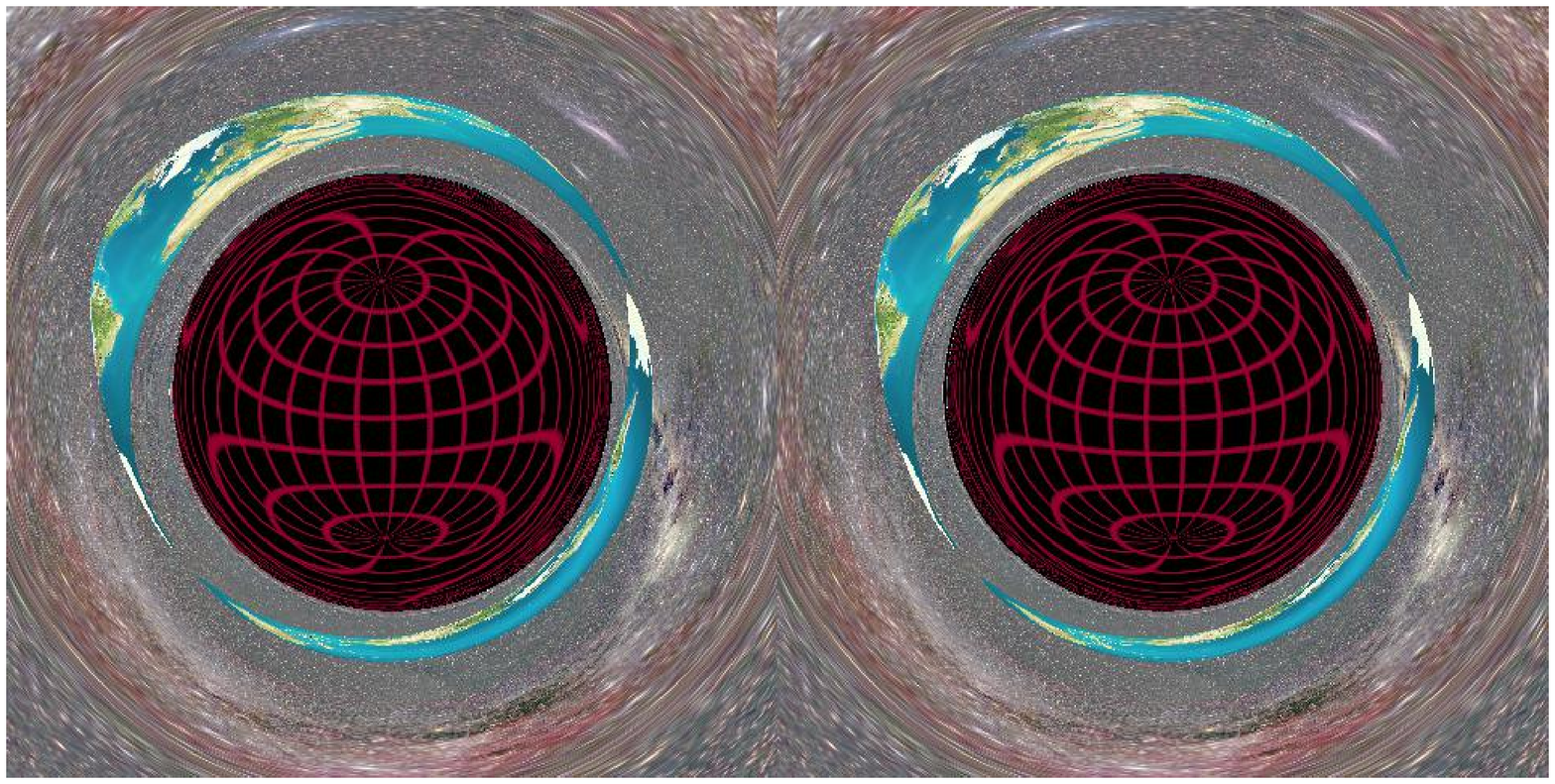}
    \vskip.5ex
    \includegraphics[width=4.5in]{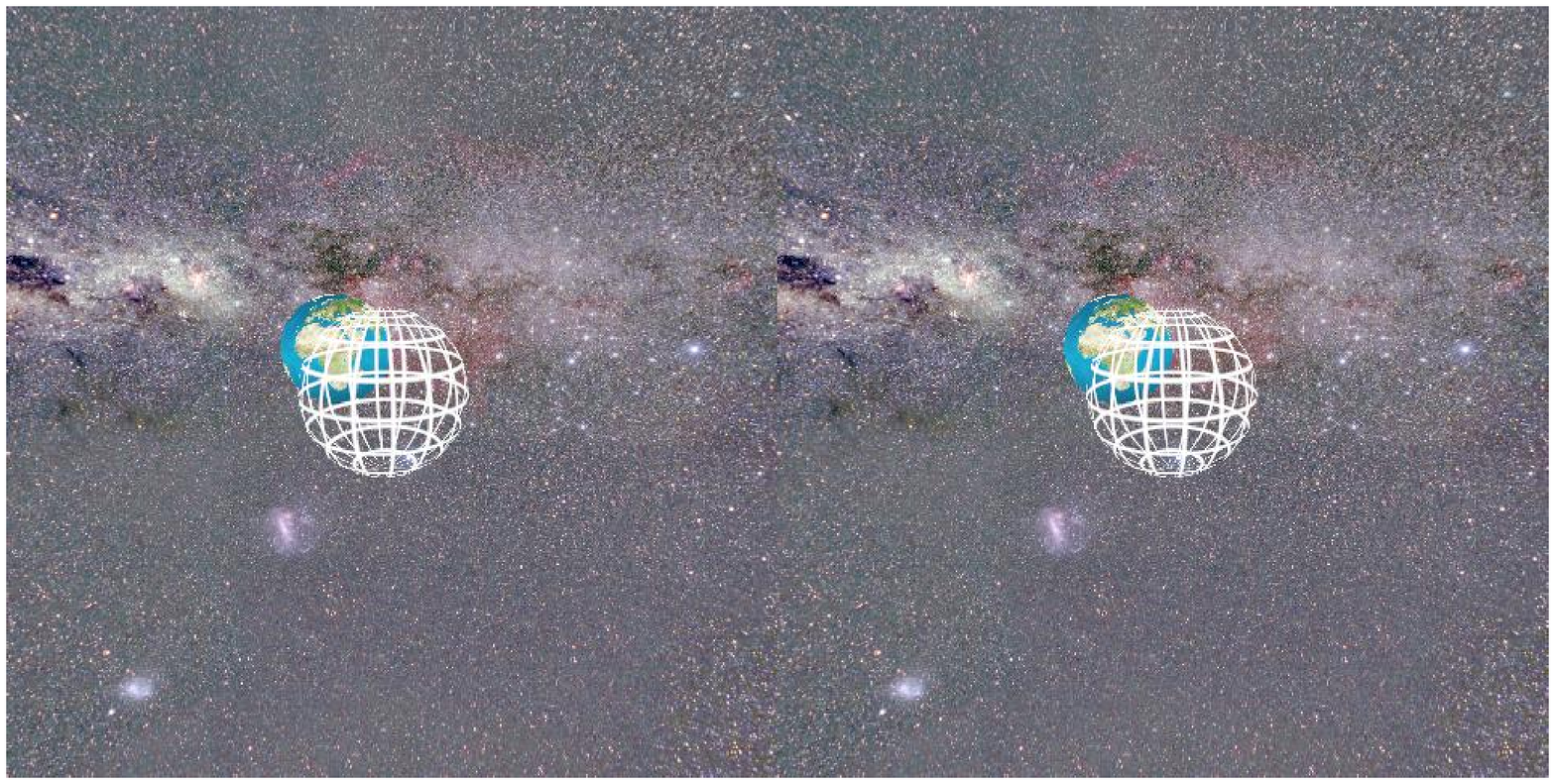}
    \caption{
(Top) the binocular view,
compared to
(middle) the stereo view using the affine distance,
and
(bottom) the binocular/stereo view with lensing turned off,
of the Earth behind a Schwarzschild black hole,
observed by an observer at rest
from a distance of
5 Schwarzschild radii.
The binocular view
at top is confusing
because the image pair
fails to mesh in the fashion that binocular vision expects.
The background is Axel Mellinger's Milky Way
\cite{Mellinger}
(with permission).
To view, cross your eyes, and relax your focus.
    }
    \label{earthbinocular}
    \end{figure}
}

\newcommand{\beamfig}{
    \begin{figure}
    \centering
    \includegraphics[scale=.9]{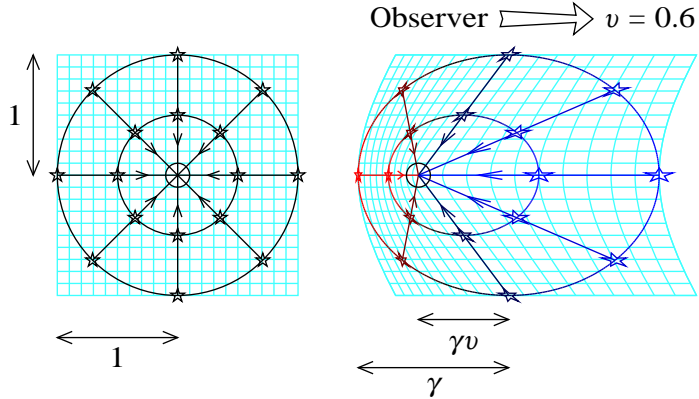}
    \caption{
In special relativity,
the scene seen by an observer moving through the scene (right)
is relativistically beamed
compared to the scene seen by an observer at rest relative to the scene (left).
On the left,
the observer
at the center of the circle is at rest
relative to the surrounding scene.
On the right,
the observer
is moving to the right through
the same scene at $v = 0.6$ times the speed of light.
The arrowed lines represent energy-momenta of photons.
The length of an arrowed line
is proportional to the affine distance,
which is proportional to the perceived energy of the photon.
The scene ahead of the moving observer appears concentrated, blueshifted,
and farther away,
while the scene behind appears expanded, redshifted, and closer.
    }
    \label{beam}
    \end{figure}
}

\newcommand{\coordfig}{
    \begin{figure}
    \centering
    \includegraphics[scale=.7]{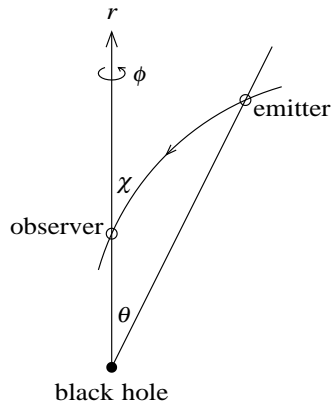}
    \caption{
Coordinate system.
The emitter and observer are separated by angle $\theta$
in a polar coordinate system
$\{ r , \theta, \phi \}$
about the black hole.
A light ray from emitter to observer
subtends apparent angle
$\chi$
from the vertical axis.
   }
    \label{coord}
    \end{figure}
}

\newcommand{\sceneoneframesfig}{
    \begin{figure}
    \centering
    \includegraphics[width= 2.5in]{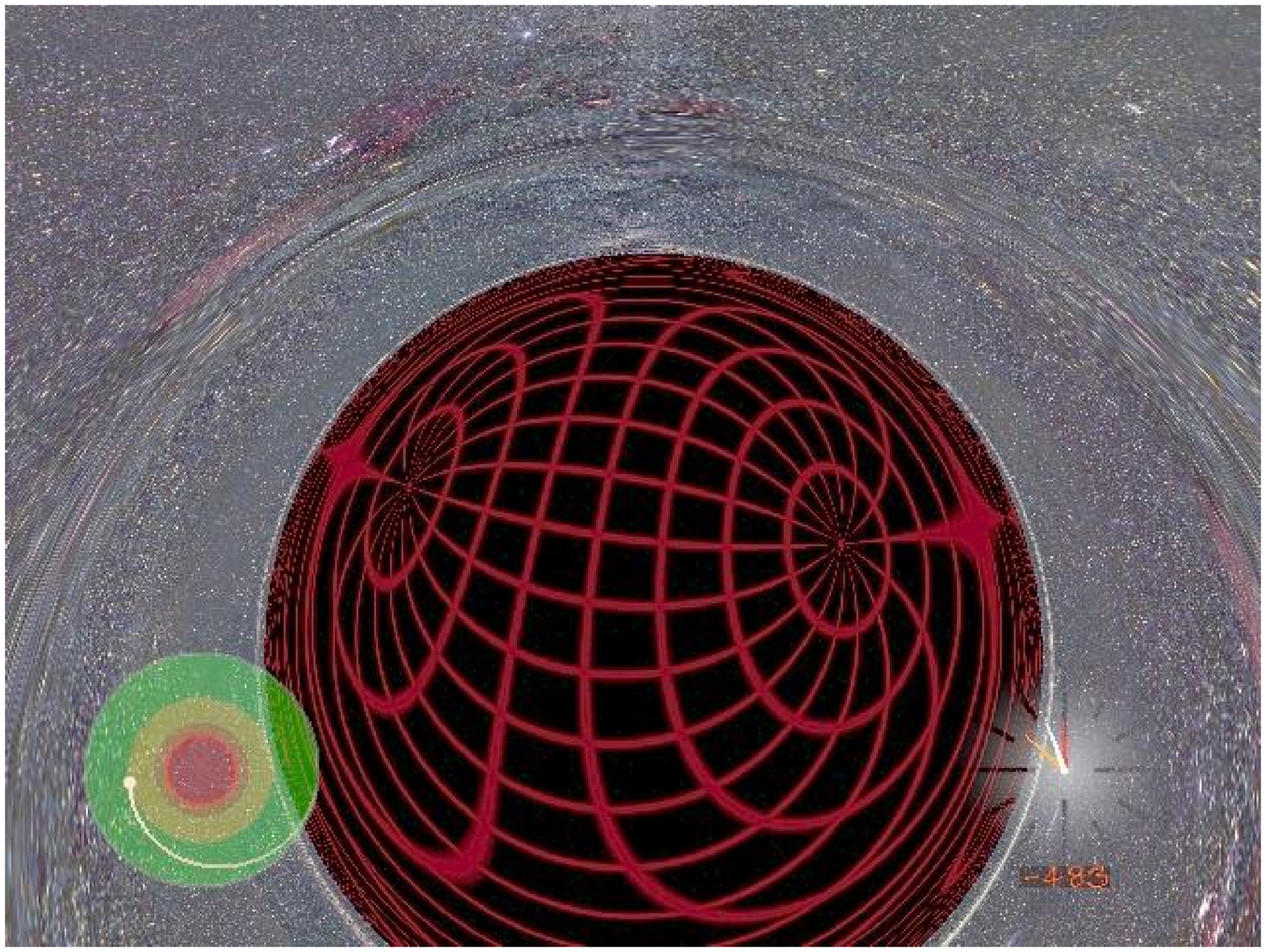}
    \includegraphics[width= 2.5in]{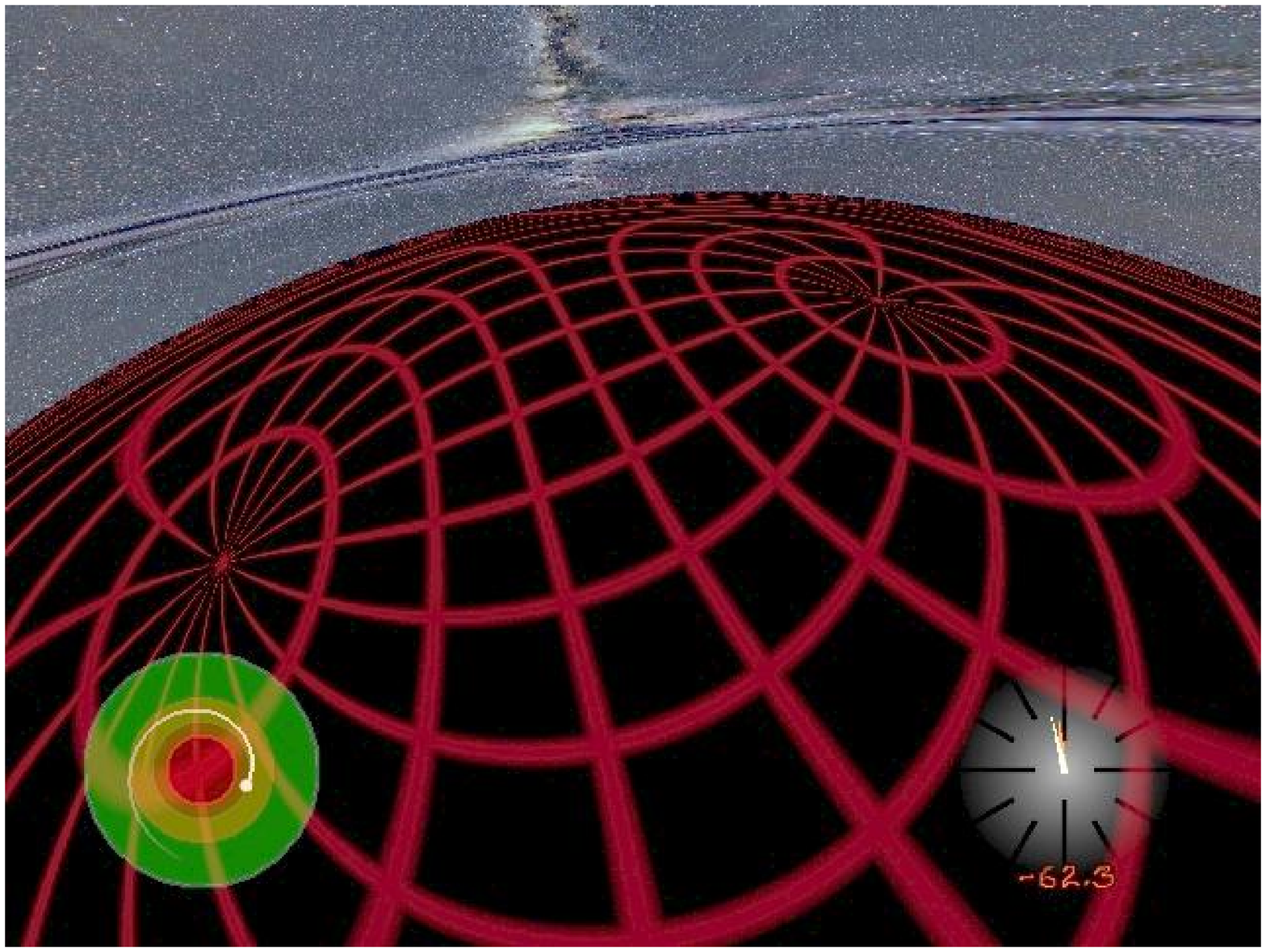}
    \vskip.5ex
    \includegraphics[width= 2.5in]{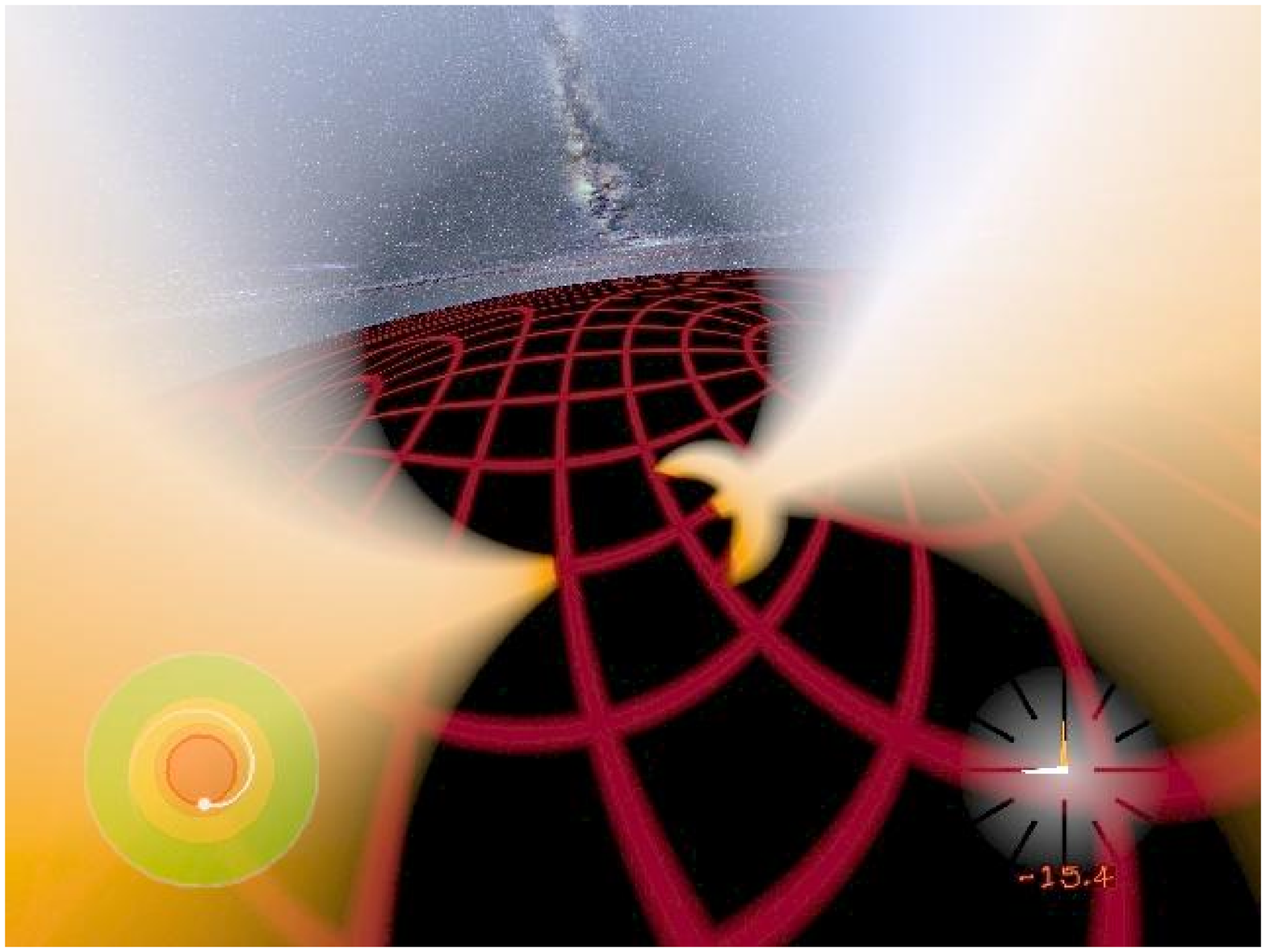}
    \includegraphics[width= 2.5in]{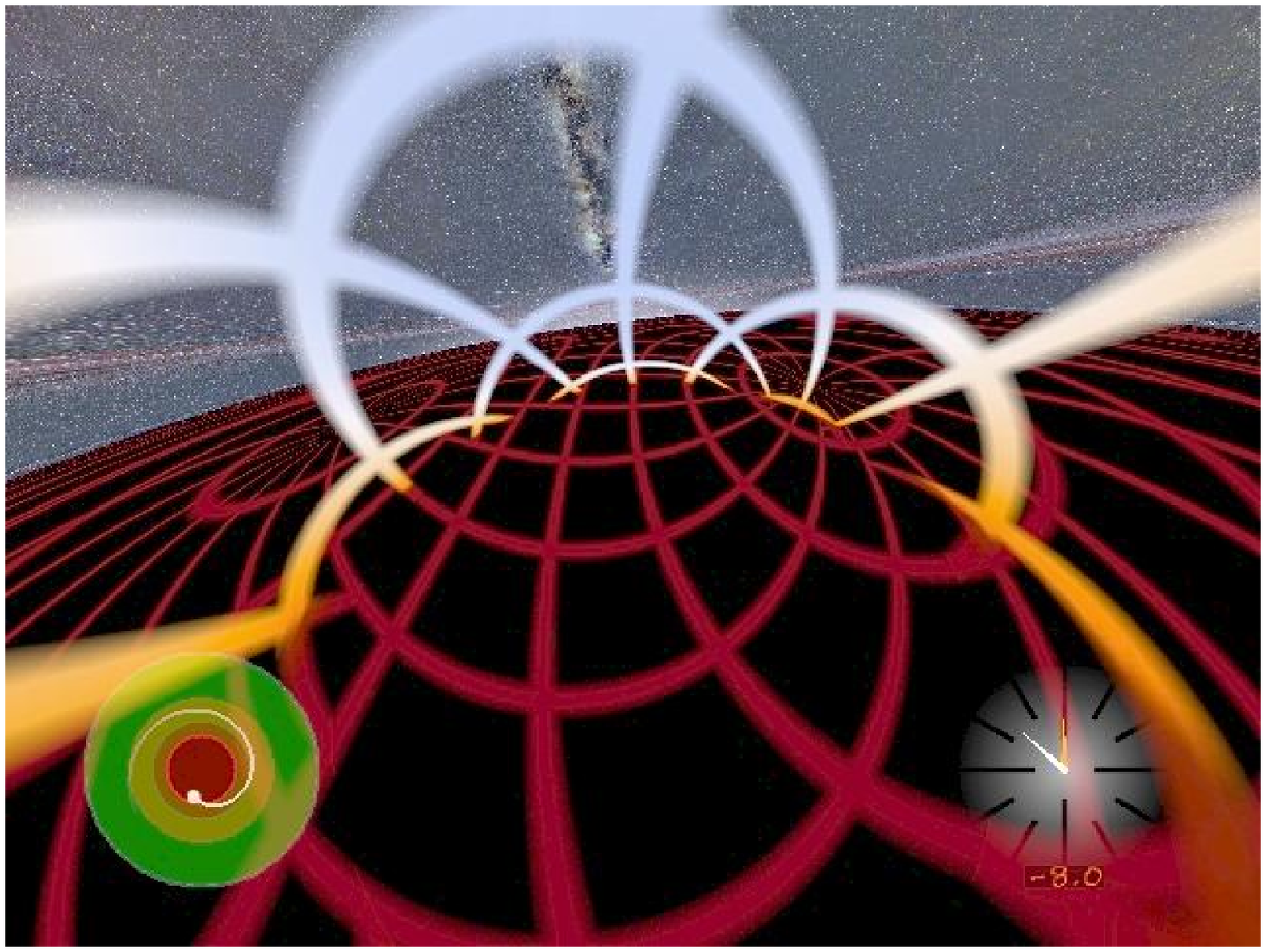}
    \vskip.5ex
    \includegraphics[width= 2.5in]{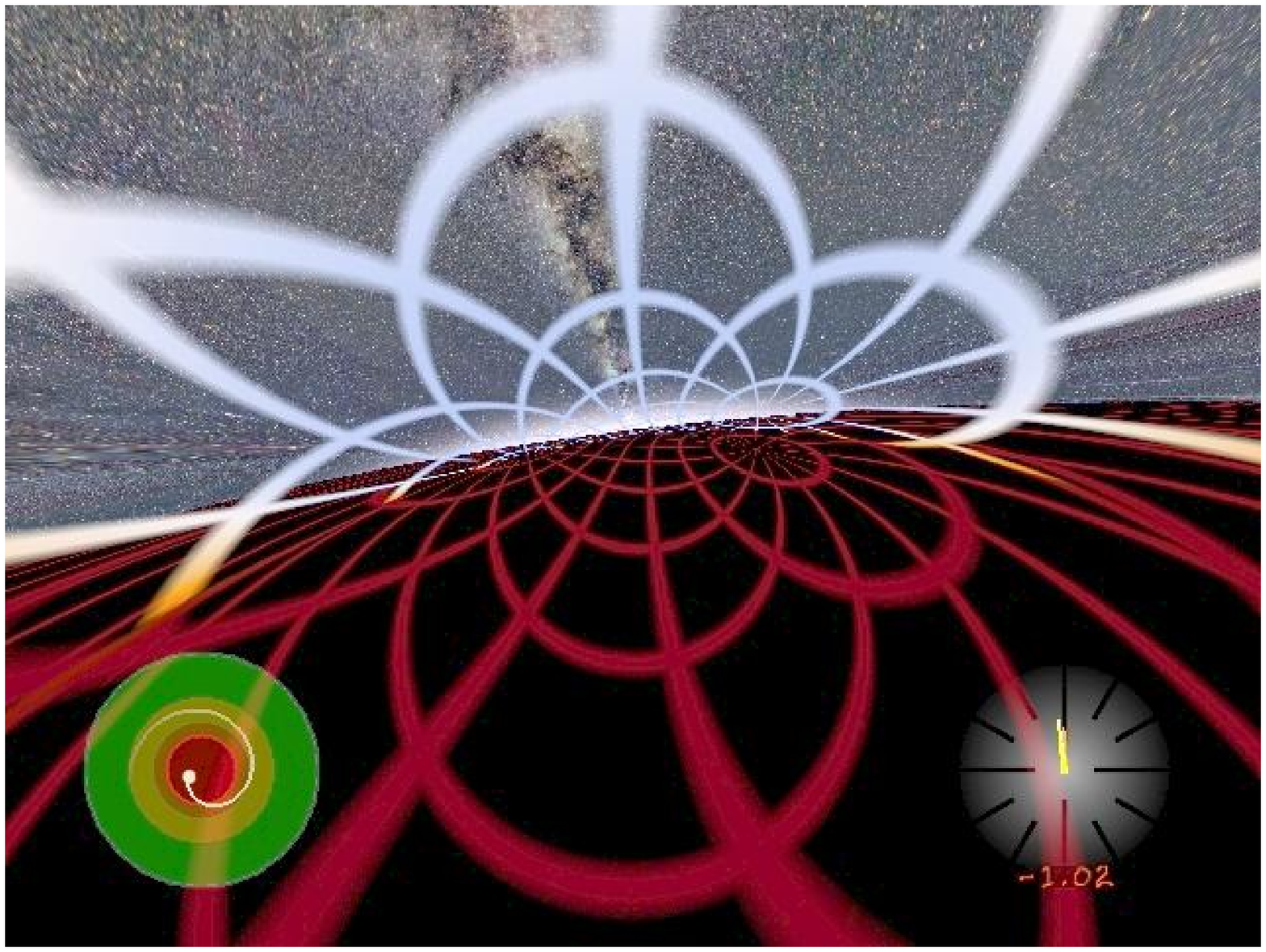}
    \includegraphics[width= 2.5in]{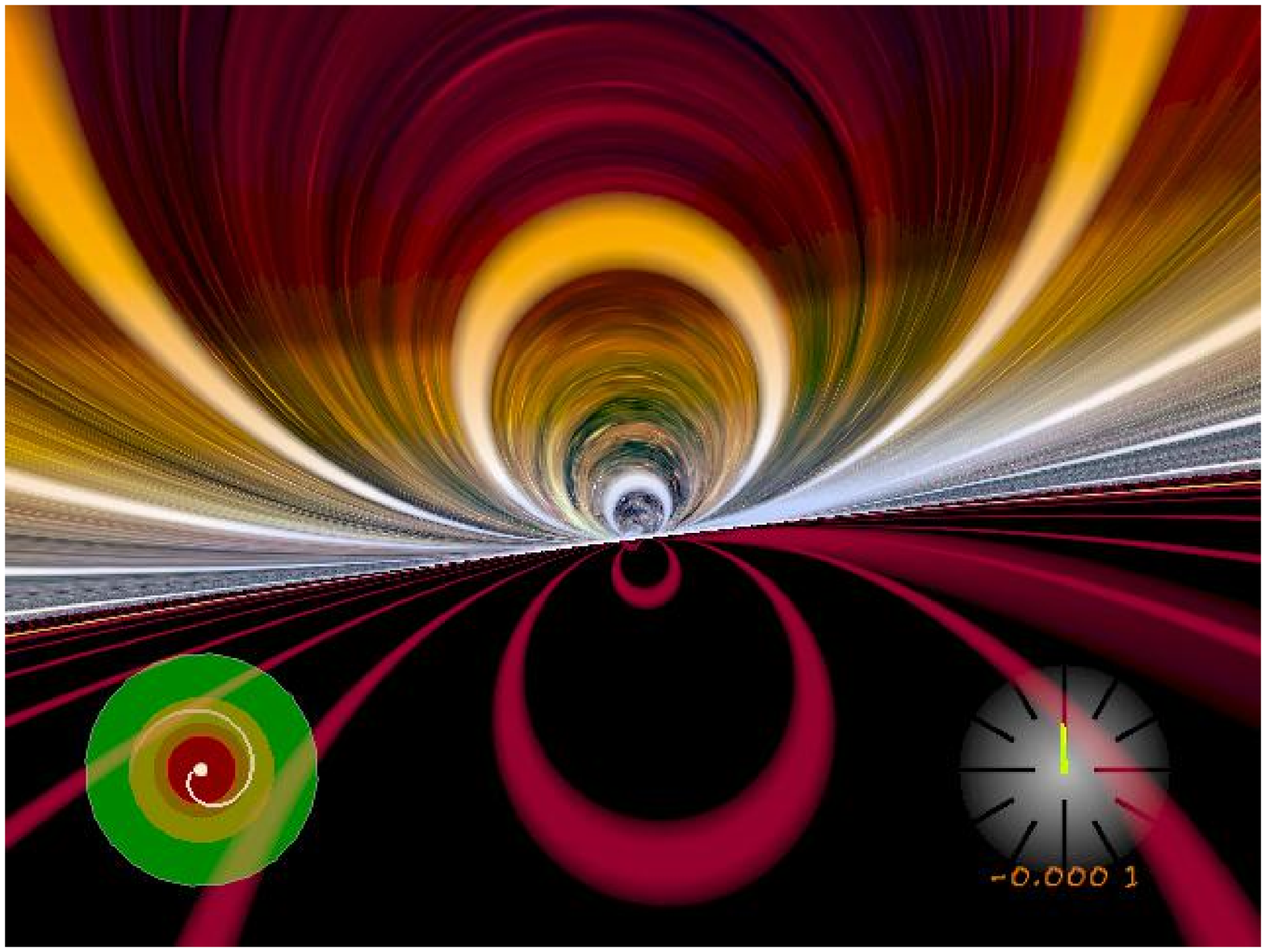}
    \caption{
Six frames from a visualization
\protect\cite{Hamilton:website}
of the scene
seen by an observer falling into a Schwarzschild black hole
on a geodesic with specific energy and angular momentum
$E = 1$ and $L = 3.92 M$
geometric units.
From left to right and top to bottom,
the observer is at radii of
$6.011 M$,
$3.000 M$,
$1.997 M$,
$1.613 M$,
$0.786 M$,
and
$0.045 M$
geometric units.
The illusory horizon is painted with a dark red grid,
while the true horizon is painted with a grid
colored with an appropriately red- or blue-shifted blackbody color.
The schematic map at the lower left of each frame
shows the trajectory (white line) of the observer through regions
of stable circular orbits (green), unstable circular orbits (yellow),
no circular orbits (orange), the horizon (red line),
and inside the horizon (red).
The clock at the lower right of each frame
shows the proper time left to hit the singularity,
in seconds, scaled to a black hole mass of $5 \times 10^6 \unit{\Msun}$,
similar to the mass of the Milky Way's supermassive black hole
\protect\cite{Ghez:2003qj,Eisenhauer:2005cv}.
The background is Axel Mellinger's Milky Way
\cite{Mellinger}
(with permission).
    }
    \label{scene1frames}
    \end{figure}
}

\newcommand{\bhholoschwfig}{
    \begin{figure}
    \centering
    \includegraphics[scale=1]{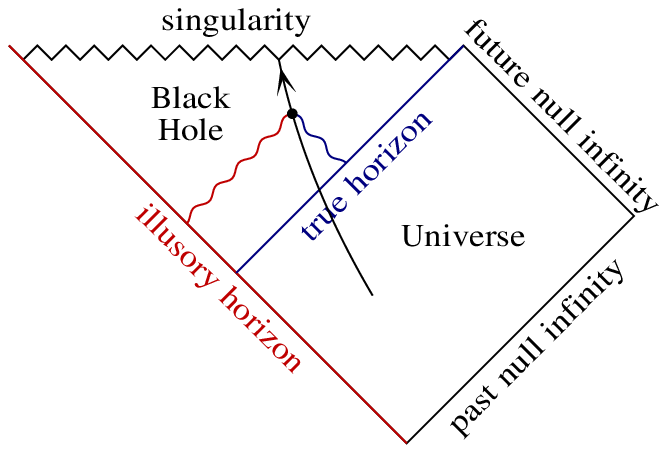}
    \caption{
Penrose diagram of a Schwarzschild black hole.
The arrowed line represents the worldline of an infaller.
The wiggly lines represent ingoing (blue) and outgoing (red) photons
emitted from the true and illusory horizons.
In the analytically extended Schwarzschild geometry,
the illusory horizon is a true horizon,
the horizon of a white hole and parallel universe.
In a real black hole formed from the collapse of a star,
the illusory horizon is replaced by the exponentially redshifting
and dimming surface of the collapsing star.
As time goes by,
the exponentially dimming surface becomes indistinguishable
from the white hole horizon of the extended Schwarzschild geometry.
In Figures~\ref{scene1frames} and \ref{scenestereo},
the black disk painted with a dark red grid is the illusory horizon,
while the whitish grid is the true horizon.
    }
    \label{bhholo_schw}
    \end{figure}
}

\newcommand{\scenestereofig}{
    \begin{figure*}
    \centering
    \includegraphics[width=3.7in]{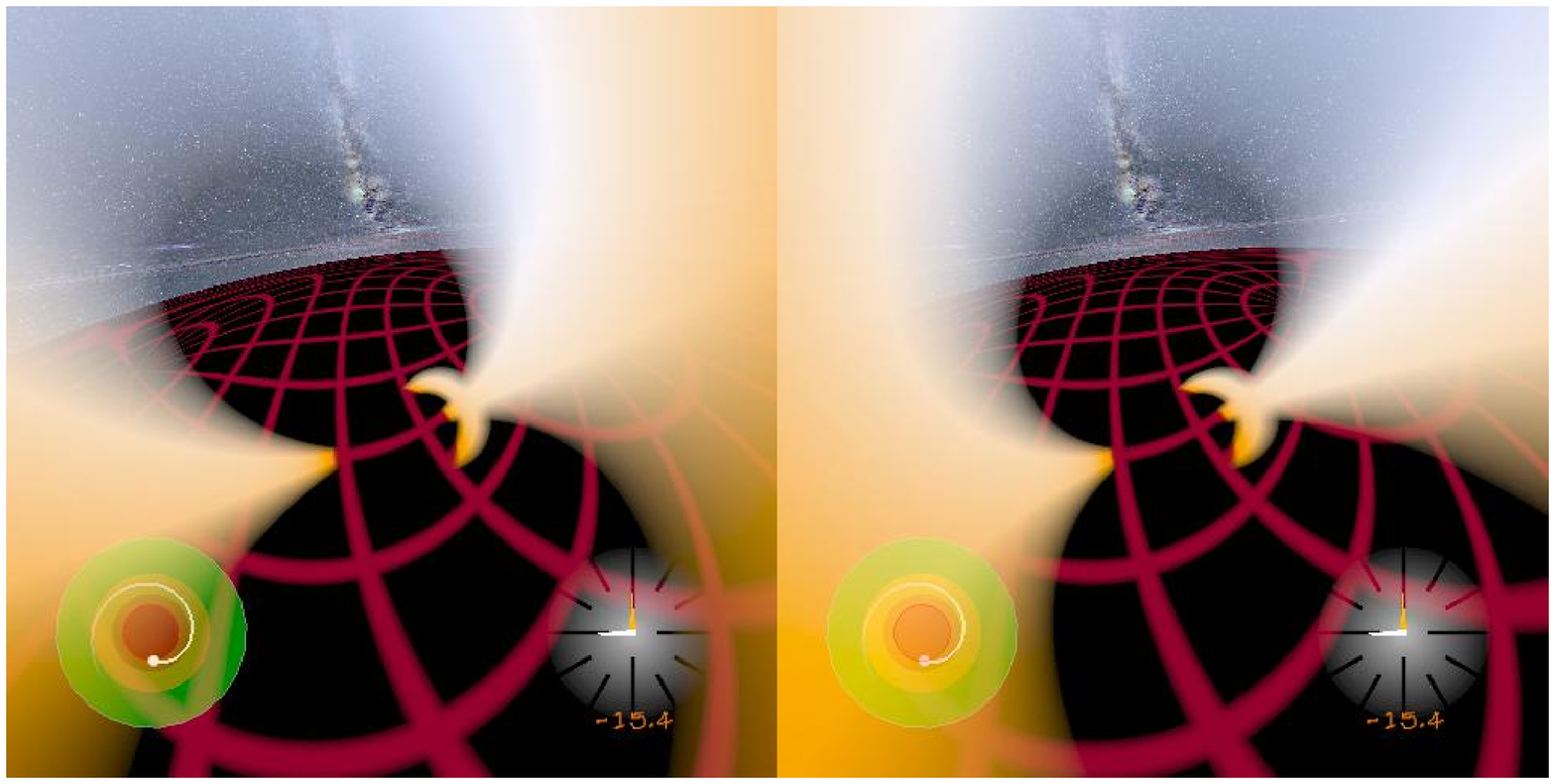}
    \vskip.5ex
    \includegraphics[width=3.7in]{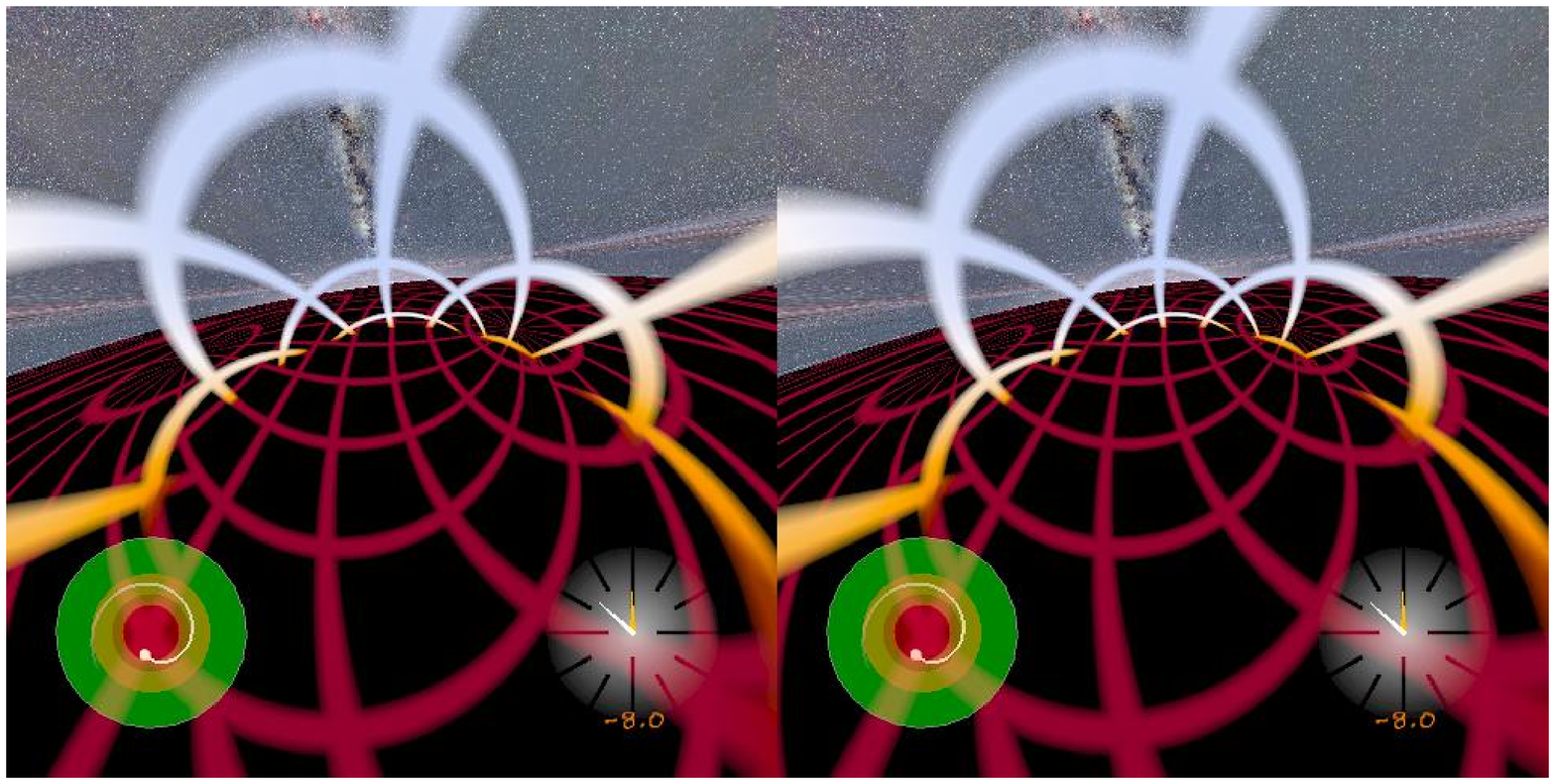}
    \vskip.5ex
    \includegraphics[width=3.7in]{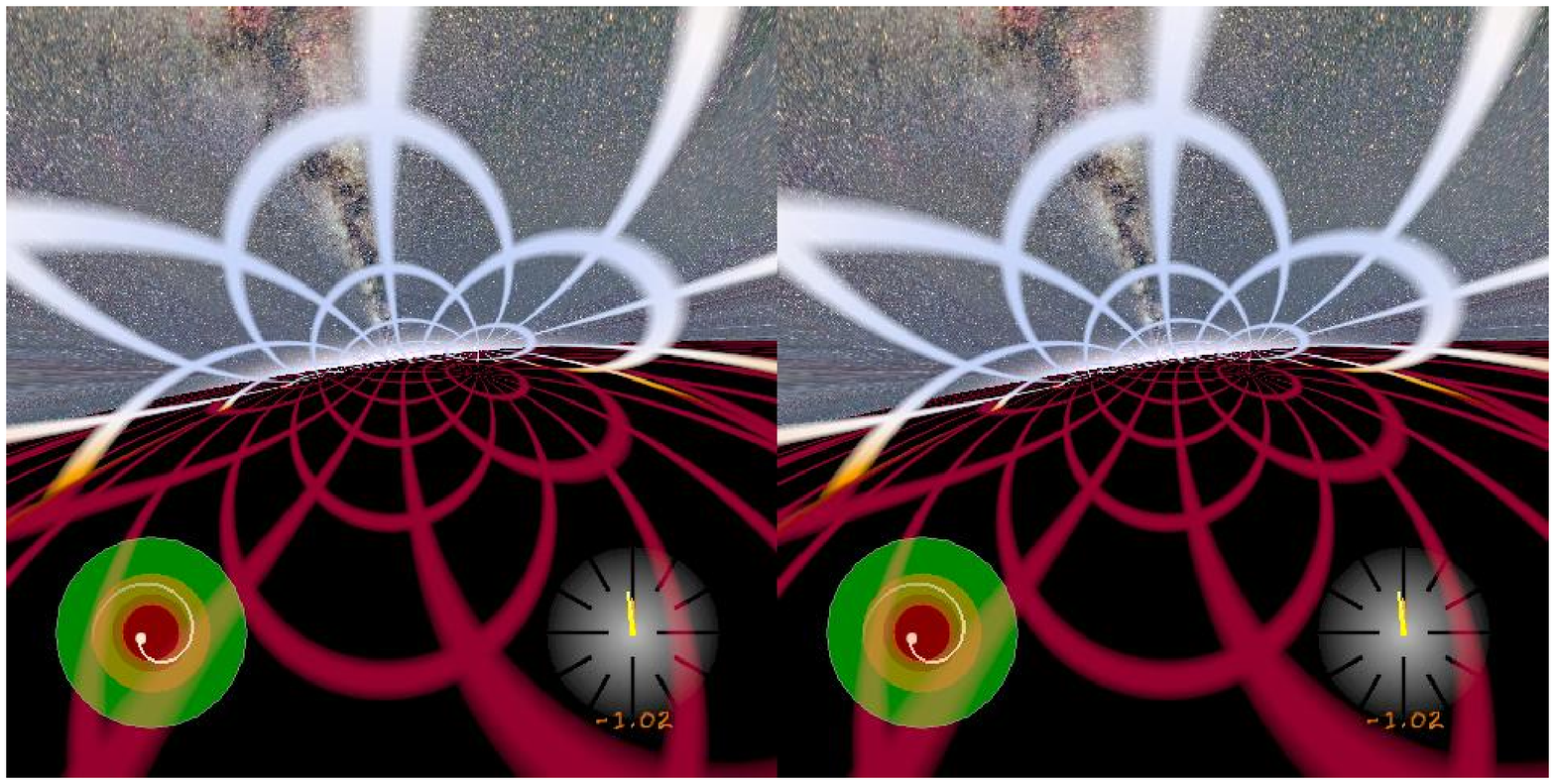}
    \vskip.5ex
    \includegraphics[width=3.7in]{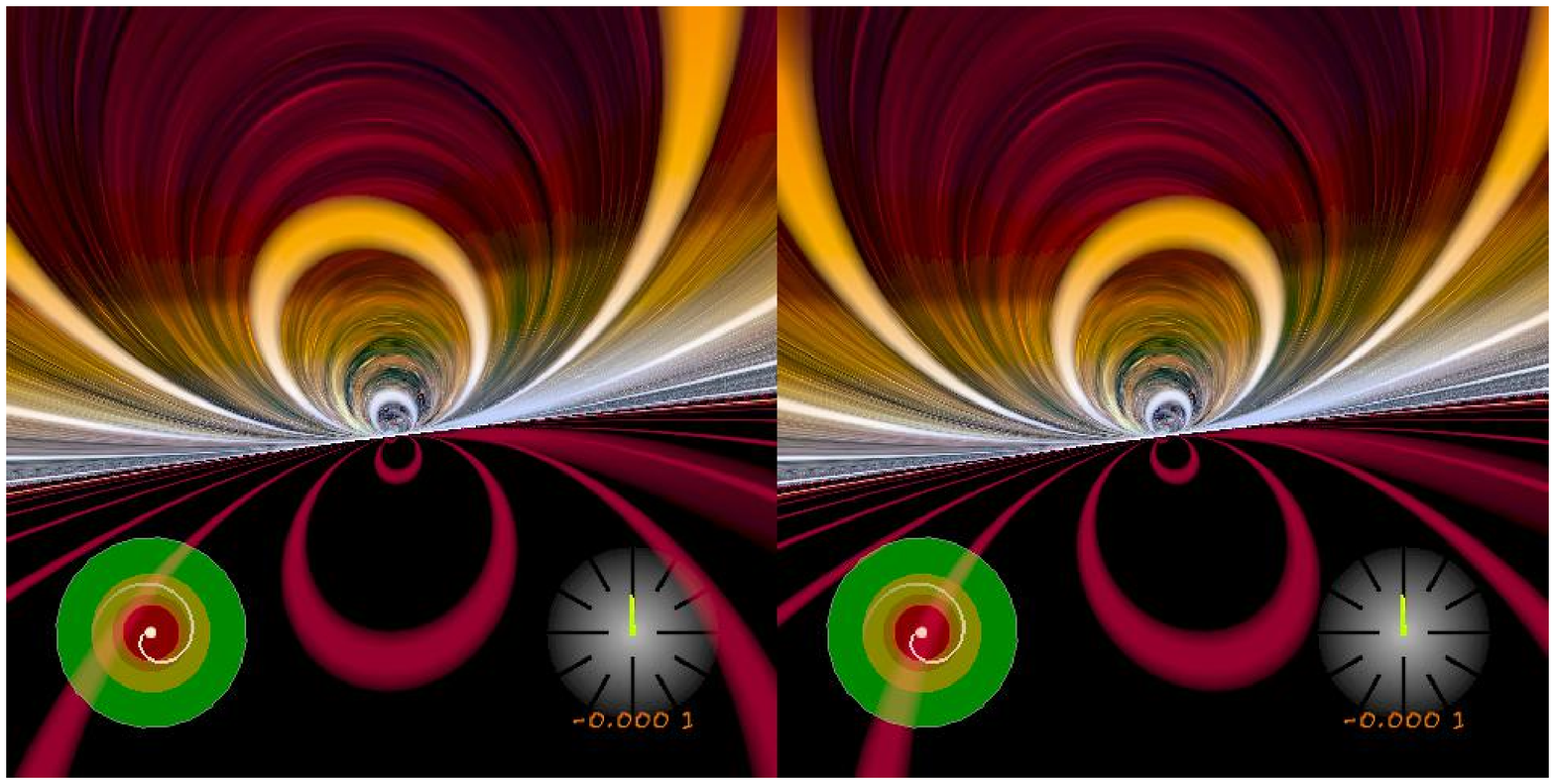}
    \caption{
Stereo versions of frames
three to six
of Figure~\protect\ref{scene1frames},
with the observer at radii of
$1.997 M$,
$1.613 M$,
$0.786 M$,
and
$0.045 M$
geometric units.
To view, cross your eyes, and relax your focus.
The horizon grids should pop out,
while the rest of the scene remains in the background.
    }
    \label{scenestereo}
    \end{figure*}
}

\newcommand{\sceneonefig}{
    \begin{figure}[t!]
    \centering
    \includegraphics[scale=.95]{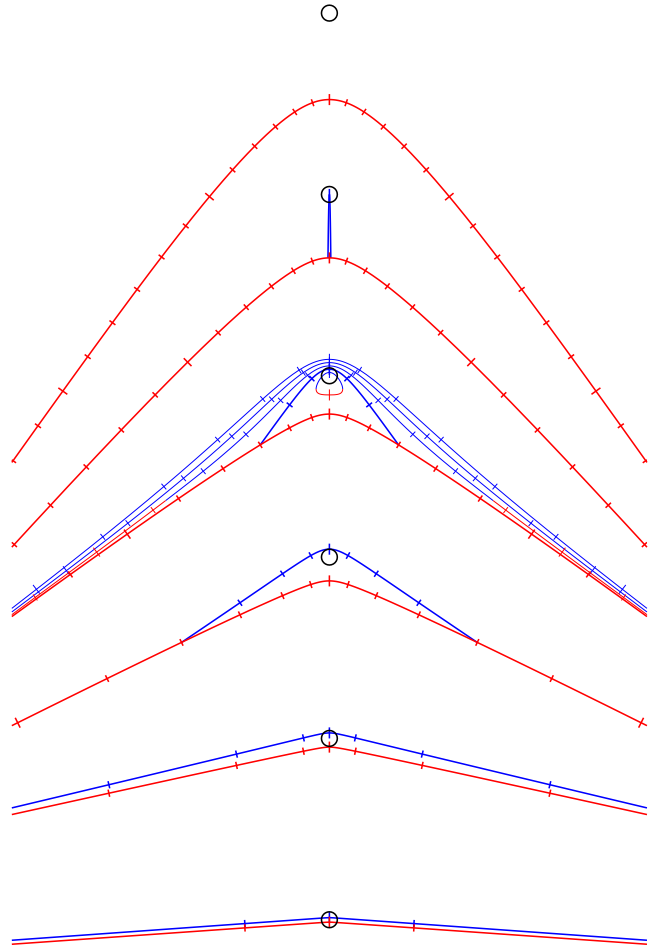}
    \caption{
Cross-sections through the effective 3-dimensional scene
perceived by an observer
who free-falls radially into the Schwarzschild black hole
from zero velocity at infinity,
with $E = 1$ and $L = 0$.
The six successive views show the
location of the true (blue) and illusory (red) horizons
From top to bottom,
the observer
is at radii of
$3 M$, $1.999 M$, $1 M$, $0.5 M$, $0.1 M$, and $0.01 M$ geometric units.
The small circle, diameter $1 M$ geometric unit,
at the center of each frame
marks the position of the observer.
The third-from-top frame,
where the observer is at $1 M$ geometric unit,
shows additional detail:
the various lines show the perceived location of emitting surfaces at,
from top to bottom,
radii of
$3.5 M$, $3 M$ (photon sphere), $2.5 M$, $2 M$ (horizon), and $1.5 M$ geometric units.
Tick marks on the horizons are spaced every $30^\circ$.
The lines are blue where the emitted photon is initially ingoing,
and red where the emitted photon is initially outgoing.
    }
    \label{scene1}
    \end{figure}
}

\newcommand{\schwonestereofig}{
    \begin{figure}[t]
    \centering
    \includegraphics[width=5in]{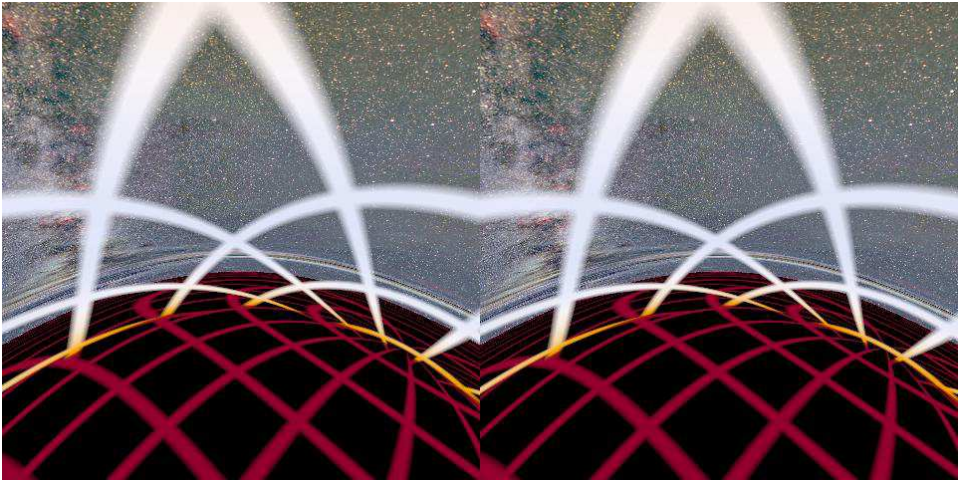}
    \caption{
Stereo version of
the scene seen by an observer free-falling radially
from zero velocity at infinity,
with specific energy $E = 1$ and angular momentum $L = 0$.
The observer is at a radius of $1 M$~geometric unit,
as in the third-from-top frame of Figure~\protect\ref{scene1}.
The observer is looking in a direction $30^\circ$ below horizontal.
The grid above is the true horizon,
while the grid below is the illusory horizon.
    }
    \label{schw1stereo}
    \end{figure}
}


\newcommand{\bhholoschwzerofig}{
    \begin{figure}
    \centering
    \includegraphics[scale=1]{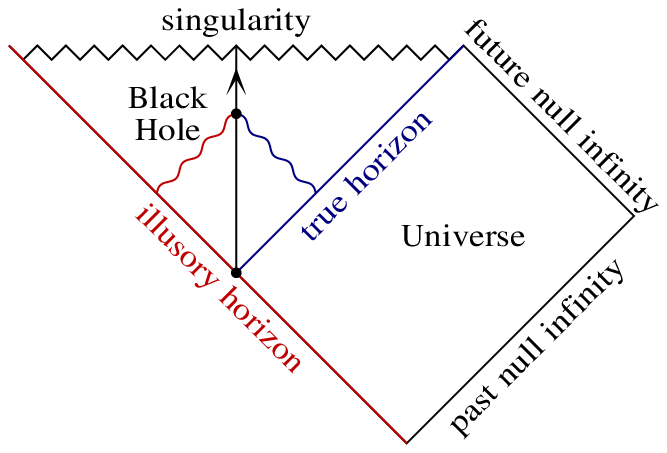}
    \caption{
Penrose diagram of the Schwarzschild black hole,
with the vertical arrowed line showing the worldline of
an infaller who falls on the zero-energy radial geodesic.
The wiggly lines represent ingoing (blue) and outgoing (red) photons
emitted from the true and illusory horizons.
    }
    \label{bhholo_schw0}
    \end{figure}
}

\newcommand{\scenefig}{
    \begin{figure}
    \centering
    \includegraphics[scale=.95]{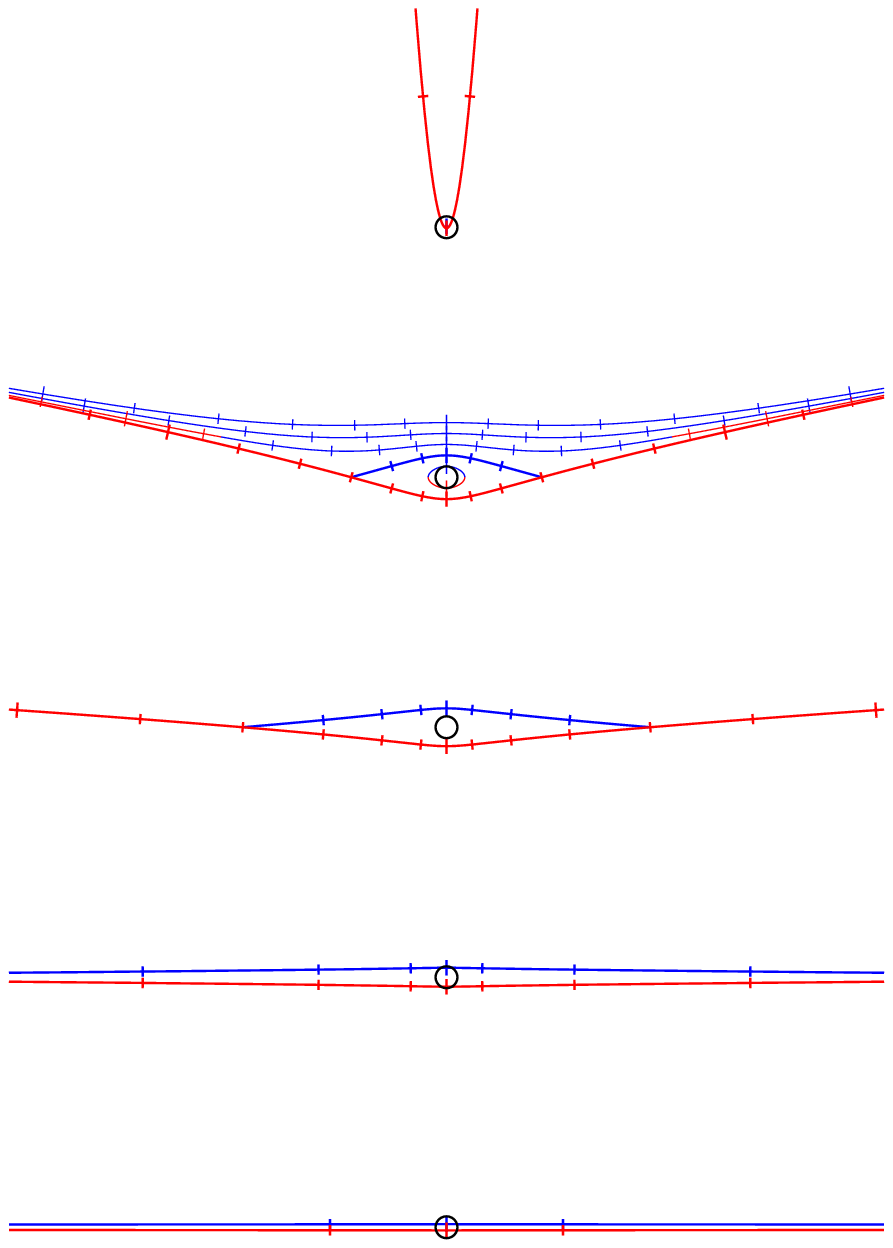}
    \caption{
Cross-sections through the effective 3-dimensional scene,
similar to Figure~\protect\ref{scene1},
but as perceived by an observer
who is infalling on the zero-energy radial geodesic,
which makes manifest the symmetry between ingoing and outgoing.
The zero-energy geodesic exists only inside the horizon,
so only images where the observer is inside the horizon can be shown.
From top to bottom,
the observer is at radii of
$1.999 M$, $1 M$, $0.5 M$, $0.1 M$, and $0.01 M$ geometric units,
as in all but the top frame of Figure~\protect\ref{scene1}.
The second-from-top panel,
where the observer is at $1 M$ geometric unit,
shows additional detail
similar to the third-from-top panel of Figure~\ref{scene1}:
the various lines show the perceived location of emitting surfaces at,
from top to bottom,
radii of
$3.5 M$, $3 M$ (photon sphere), $2.5 M$, $2 M$ (horizon), and $1.5 M$ geometric units.
The lines are blue where the emitted photon is initially ingoing,
and red where the emitted photon is initially outgoing.
    }
    \label{scene}
    \end{figure}
}

\newcommand{\schwzerostereofig}{
    \begin{figure}
    \centering
    \includegraphics[width=5in]{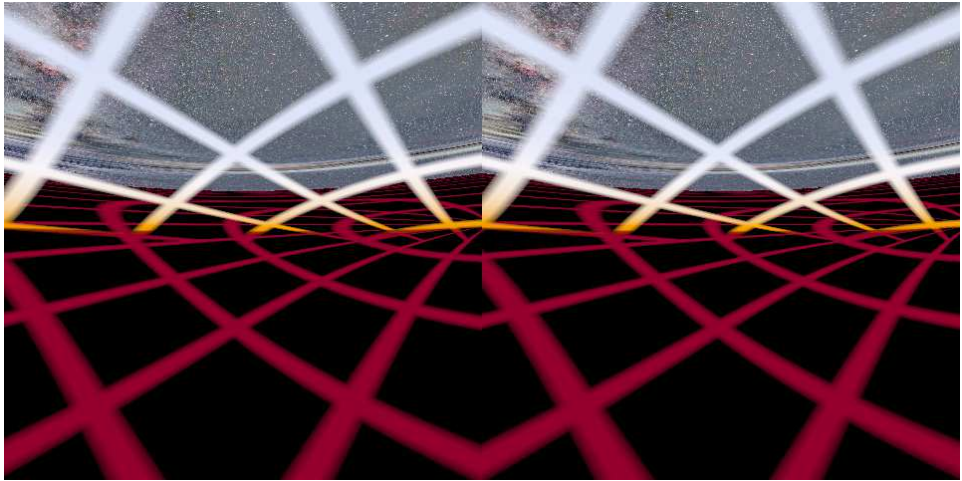}
    \caption{
Stereo version of
the scene seen by an observer free-falling radially on the
zero energy geodesic, $E = L = 0$.
The observer is at a radius of $1 M$~geometric unit,
as in the second-from-top frame of Figure~\protect\ref{scene}.
The observer is looking in the horizontal direction.
The scene is the same as in Figure~\protect\ref{schw1stereo},
Lorentz-boosted radially upward.
    }
    \label{schw0stereo}
    \end{figure}
}

\newcommand{\magfig}{
    \begin{figure}[h]
    \centering
    \includegraphics[scale=.9]{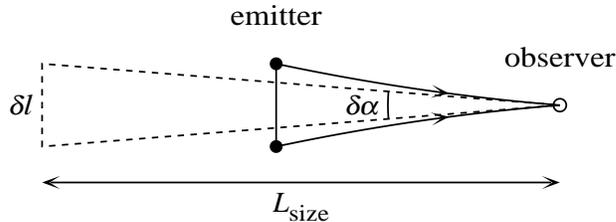}
    \caption{
The size distance $L_{\rm size}$
follows from the
apparent angular size
$\delta \alpha$
of an object
of known proper size
$\delta l$.
    }
    \label{mag}
    \end{figure}
}

\newcommand{\distfig}{
    \begin{figure}[t!]
    \centering
    \includegraphics[scale=.8]{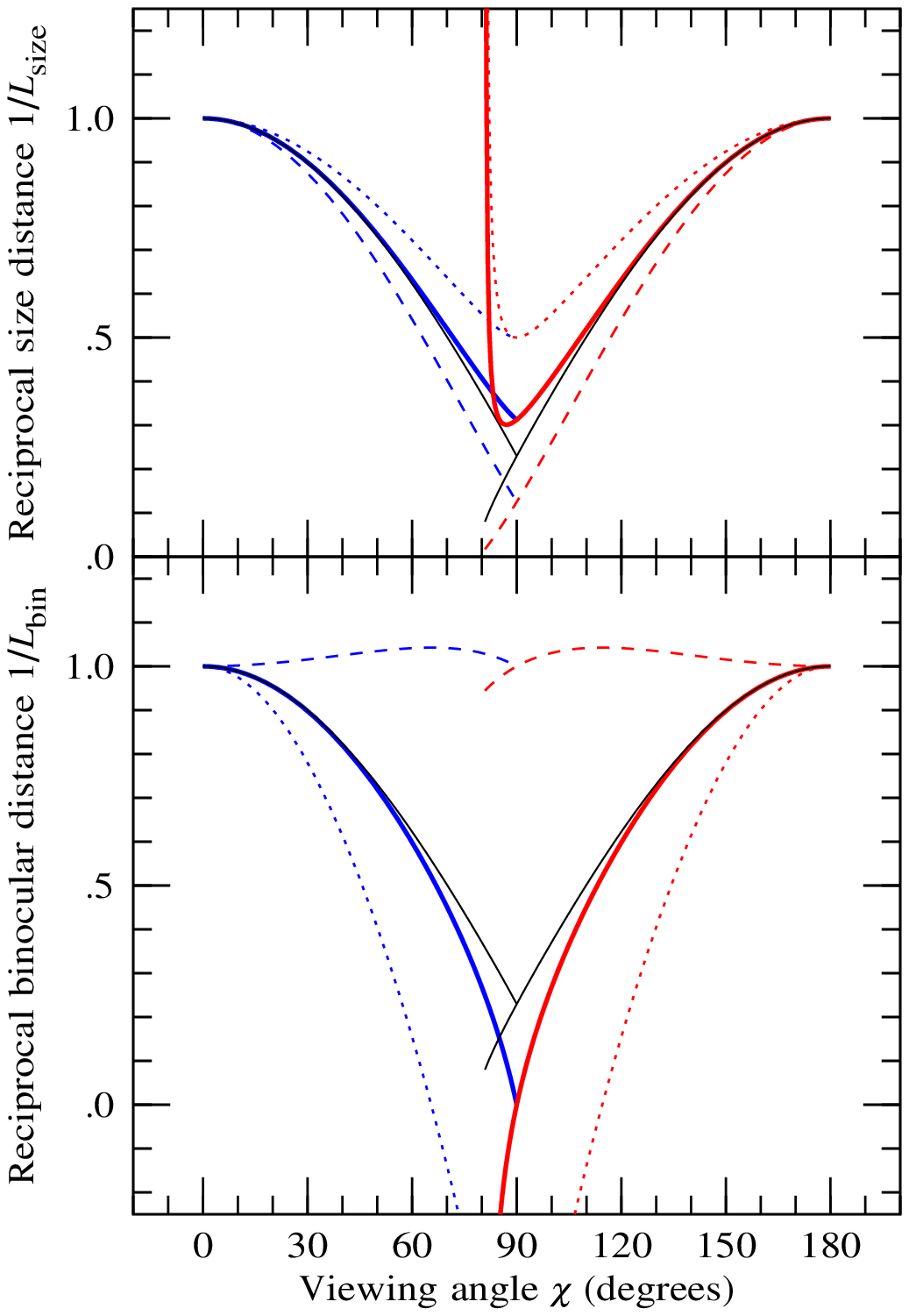}
    \hskip.1in
    \includegraphics[bb=250 219 397 648,scale=.8]{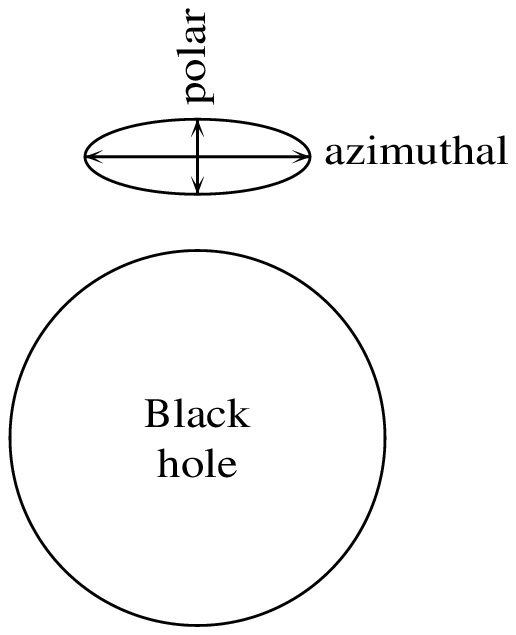}
    \caption{
Perceived
reciprocal
size distance
$1/L_{\rm size}$
(upper panel)
and
binocular distance
$1/L_{\rm bin}$
(lower panel)
to the true (blue) and illusory (red) horizons
of a Schwarzschild black hole
perceived by an observer inside the black hole,
as a function of the viewing angle $\chi$ relative to the vertical axis,
Figure~\protect\ref{coord},
with $0^\circ$ being up to the sky above,
and $180^\circ$ down to the black hole below.
The observer
is at a radius of 1 geometric unit,
and
is infalling on the zero-energy radial geodesic,
the same as illustrated in
Figure~\protect\ref{schw0stereo}
and the second-from-top panel of
Figure~\protect\ref{scene}.
Dashed and dotted lines show the reciprocal distances
respectively in the polar and azimuthal directions
(per the diagram at right),
while thick solid lines show the average of the polar and azimuthal
reciprocal distances.
The averaged reciprocal distances
provide a good estimate of the affine distance
(thin solid black lines)
for viewing angles not too far off axis.
    }
    \label{dist}
    \end{figure}
}

\newcommand{\binfig}{
    \begin{figure}[h]
    \centering
    \includegraphics[scale=.85]{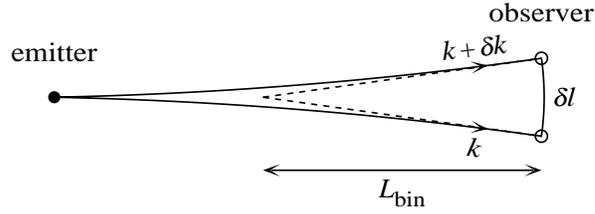}
    \caption{
The binocular distance $L_{\rm bin}$
is the radius of curvature of a wavefront of light
emitted by an emitter and observed by an observer.
    }
    \label{bin}
    \end{figure}
}

\newcommand{\apefig}{
    \begin{figure}
    \centering
    \includegraphics[width=3.5in]{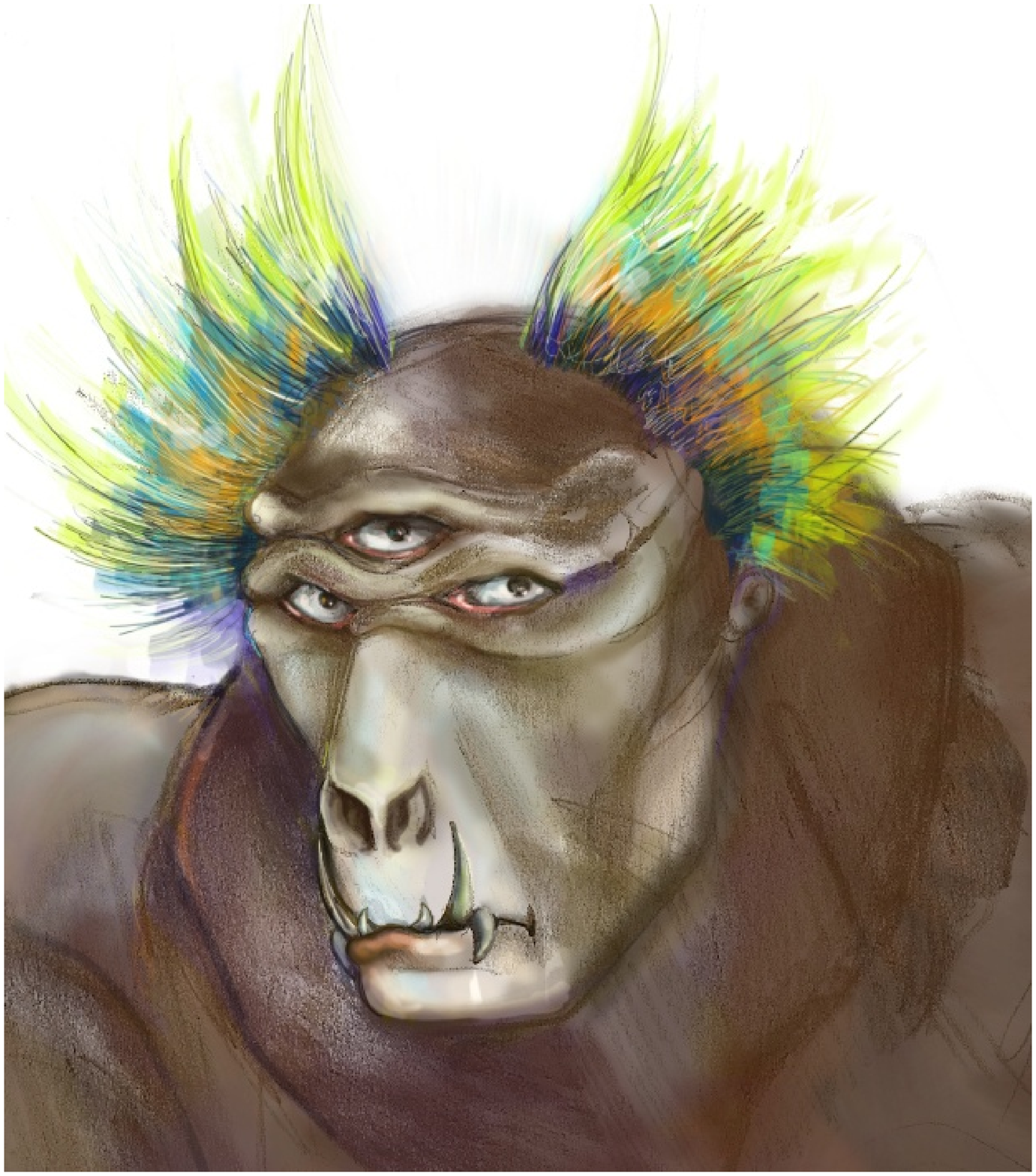}
    \caption{
An imaginative artistic rendering of
a three-eyed inhabitant of a highly curved spacetime.
The triangular configuration of eyes
would be suitable for trinocular vision.
Drawing by Wildrose Hamilton.
    }
    \label{ape}
    \end{figure}
}


\begin{abstract}
Stereoscopic visualization adds an additional dimension to the viewer's
experience, giving them a sense of distance.
In a general relativistic visualization,
distance can be measured in a variety of ways.
We argue that the affine distance,
which matches the usual notion of distance in flat spacetime,
is a natural distance to use in curved spacetime.
As an example, we apply affine distance to the visualization
of the interior of a black hole.
Affine distance is \emph{not} the distance perceived with normal binocular vision in curved spacetime.
However, the failure of binocular vision is simply a limitation of animals who have evolved in flat spacetime, not a fundamental obstacle to depth perception in curved spacetime.
Trinocular vision would provide superior depth perception.
\end{abstract}

\pacs{04.20.-q} 

\maketitle

\section{Introduction}
Visualization
has the potential to make general relativity accessible to non-experts,
and to give experts deeper insights into the theory
\cite{Ruder:2008, 10.1109/TVCG.2006.69}.
%
Presenting depth information stereoscopically%
\footnote{``Stereoscopic'' refers to image pairs that encode distance information.
In curved space-time there are various ways to measure distance,
and any one of these could be encoded in the stereoscopic view.
``Binocular'' refers specifically to the actual view seen with two eyes.}
adds an extra dimension to the viewer's experience.
The depth information has the potential to help the viewer
make better sense of a highly curved spacetime,
such as that near and inside a black hole.
As is common in relativity,
what really happens often fails to conform to expectation
drawn from slow-moving experience in flat spacetime.
Once a viewer is able to overcome preconceptions,
distance information can offer deeper insight.

As demonstrated in \S\ref{binocularfails},
normal binocular vision fails in curved spacetime.
Binocular distance measures the radius of curvature of wavefronts
of light from an emitter.
In curved spacetime, wavefronts become locally ellipsoidal
rather than spherical,
with two distinct radii of curvature in orthogonal directions.
The conflicting binocular distances would confuse a human being.
In extreme situations,
Figure~\ref{earthbinocular},
the scenes seen by two eyes a small distance apart
can differ so much as to confound binocular vision entirely.

While binocular vision fails,
stereoscopic views of curved spacetime can nevertheless be created
for human consumption
by constructing effective 3-dimensional scenes in which
objects are placed at their apparent angular positions
and at suitably chosen distances.
In \S\ref{affine} we argue that the most natural measure of
distance in general relativistic visualization is the affine distance.
The affine distance coincides with binocular distance in flat spacetime,
but differs in curved spacetime.

In \S\ref{sceneinside} we use affine distance
to construct stereoscopic visualizations near and inside
a Schwarzschild black hole.

The affine distance measures the distance to where an object used to be
at the time it emitted its light.
This distance is not directly measurable by an observer using only
local measurements,
no more than an astronomer can reach out and touch a quasar.
In \S\ref{perceptual} we discuss perceptual distances
that an observer could use to estimate the affine distance.
While binocular vision does not give the viewer a consistent sense of distance in curved spacetime,
trinocular vision performs better,
as discussed in \S\ref{trinocular}.

All the visualizations in this paper were created with the
Black Hole Flight Simulator
\cite{Hamilton:bhfs}.

\lensearthbinocularfig

\section{Binocular vision fails in curved spacetime}
\label{binocularfails}


Figure~\ref{earthbinocular}
illustrates the failure of binocular vision
in curved spacetime.
The Figure shows the Earth 
behind a Schwarzschild
black hole
\cite{Hamilton:website}.

The top panel of
Figure~\ref{earthbinocular}
shows the binocular view,
as seen by two eyes a modest distance apart.
In normal flat space
the images of distant objects seen by two eyes are shifted slightly,
and the brain uses the slight difference to infer a distance.
In curved space
the images of distant objects may be altered in size and shape
enough to confound binocular vision.
In
Figure~\ref{earthbinocular},
both the Earth behind the black hole,
and the distant Galaxy,
show discordant image pairs that defeat binocular interpretation.
Whereas the right eye (left image) sees the Earth form a ring around
the black hole,
the left eye (right image) sees the Earth as two disconnected crescents.

For comparison,
the middle panel of
Figure~\ref{earthbinocular}
shows a stereo view of the same scene,
in which objects have been put at their affine distances
in an effective 3-dimensional construction of the scene,
as advocated in this paper, \S\ref{affine}.
The bottom panel shows the familiar binocular/stereo view of
the same scene with gravitational lensing by the black hole turned off.
Among other things,
gravitational lensing magnifies the apparent size of the black hole:
a distant observer sees a Schwarzschild black hole
$3 \sqrt{3} / 2 \approx 2.6$ times larger
than its Schwarzschild radius.

In the scene illustrated in Figure~\ref{earthbinocular},
the observer is at rest at a radius of 5 Schwarzshild radii
from the black hole
(a Schwarzschild radius is the radius of the black hole's horizon).
The field of view is $105^\circ$ across the diagonal
(approximately $74^\circ \times 74^\circ$).
The eyes are separated by $0.2$ Schwarzschild radii.
Grid lines on the black hole
are aligned with Galactic coordinates,
and are every $30^\circ$
in latitude and longitude.
The observer is viewing towards
Galactic longitude and latitude $l = 263.6^\circ$, $b = -16.3^\circ$.
The Earth has a radius of 1/2 a Schwarzschild radius,
and is located at a radius of 3 Schwarzschild radii from the black hole.
Details of how the Earth is placed as if it were a rigid body
in the curved spacetime are described in \ref{rigid}.
The configuration is not realistic:
if the Earth were this close to a black hole,
it would be tidally torn apart in moments.

An animated version of Figure~\ref{earthbinocular},
showing the Earth in orbit around a black hole,
is at \cite{Hamilton:website},
in both normal and stereo representations.

\section{Affine distance}
\label{affine}

The perceived distance to an object
would ideally match the distance measured with a tape measure.
In relativity however,
defining a suitable tape measure presents challenges.
The first challenge is that
the apparent length of a tape measure depends on its velocity relative to the observer.
To make measurements that are useful to the observer,
the tape measure should be at rest relative to the observer.

The second challenge is to select the correct path along which to measure the distance
to an object.
This is the path traveled by the light,
from where it left the object to where it arrives at the observer.

The third challenge is to choose the velocity of the tape measure
at all points along the light path from emitter to observer.
In the flat spacetime of special relativity,
the tape measure can be taken to be everywhere at rest
with respect to the global rest frame.
In curved spacetime, however,
there is no global rest frame.
The most natural choice is to require that,
while the tape measure necessarily moves relative to the light,
it should not \emph{accelerate} relative to the light,
that is, the light should not undergo any redshift or blueshift
as it moves from point to point along the tape measure.

In summary, the perceived distance that should be encoded in the stereoscopic visualization
is the distance traveled by the light as measured by a tape measure
that never accelerates relative to the light
and is moving with the observer's velocity at the time of observation.
While this prescription sounds complicated, it
is well-known in general relativity,
and is called the affine distance
\cite[p.~575]{MTW:1973}.

The affine distance is the affine parameter $\lambda$ normalized
in a certain way, as we will now discuss.
Mathematically,
an interval $\dd \lambda$ of affine parameter $\lambda$
along the trajectory of
a massless particle (a light ray)
is defined so that
the derivative of the coordinates $x^\mu$ along the path
with respect to affine parameter $\lambda$
equals the 4-momentum $p^\mu$
of the particle:
\begin{equation}
\label{pmu}
  {\dd x^\mu \over \dd \lambda}
  =
  p^\mu
  \ .
\end{equation}
%
The affine parameter
defined by equation~(\ref{pmu})
is defined up to an arbitrary normalization factor.
The affine distance
is defined to be the affine parameter
normalized such that it measures proper distances
in the local rest frame of the observer.
Normalizing the affine distance in this fashion
is equivalent to normalizing the affine parameter
such that the photon momentum $p^\mu$
has unit energy as perceived by the observer.
It is convenient to denote this normalized photon momentum
by the photon wavevector
$k^\mu$,
\begin{equation}
\label{kmu}
  k^\mu
  \equiv
  {\dd x^\mu \over \dd \lambda}
  \ .
\end{equation}
By construction, the wavevector is normalized to unit energy,
$k^t = E = 1$,
at the point of observation by,
and in the rest frame of,
the observer.
With this convention,
the affine distance $\lambda$ has units of length,
while the photon wavevector $k^\mu$ is dimensionless.

\beamfig

\subsection{Affine distance in special relativity}
\label{affinesr}

In the flat (Minkowski) spacetime of special relativity,
the affine distance to any object
coincides with the distance
measured with a tape measure in the observer's rest frame.
The affine distance also coincides with
both the size and binocular distances described in \S\ref{perceptual}.
Thus in flat spacetime
all the usual rules of 3D stereo rendering and perception apply.

If two observers at the same point are moving relative to each other,
then they measure a different affine distance along the same light ray.
Each observer regards the affine distance as measuring
the number of wavelengths along the path of the photon.
The number of wavelengths is proportional to the frequency, or energy,
of the photon.
Thus the affine distance $\lambda$
to an emitter measured by an observer
is proportional to the perceived energy $E$ of the photon
\begin{equation}
\label{lambdaE}
  \lambda \propto E
  \ .
\end{equation}

Figure~\ref{beam}
illustrates how this works.
On the left side of the figure,
the observer (unfilled circle)
at the center is at rest relative the surrounding scene.
The stars represent parts of the scene.
Two sets of stars are shown,
one at distance $0.5$, the other at distance $1$ from the observer.
On the right side of the figure,
the observer (unfilled circle) is moving
through the same scene
to the right at velocity
$v = 0.6$ times the speed of light.
For both stationary and moving observers,
the arrowed lines converging on the observer
represent energy-momenta of photons emitted by the stars.
The energy-momenta of photons are null 4-vectors.
The photon energy-momenta perceived by the two observers
are Lorentz-boosted relative to each other.
If, relative to the observer at rest,
the stars lie on a celestial sphere of radius $1$,
then, relative to the moving observer,
the same stars lie on a celestial ellipsoid
that is stretched by the Lorentz gamma factor $\gamma$
in the direction of motion,
and the observer is displaced
a distance $\gamma v$ from the center of the ellipsoid
to a focus of the ellipsoid.
The Lorentz boost produces the relativistic aberration
and red/blue-shifting illustrated in the figure.
The moving observer sees the scene ahead to be concentrated, blueshifted,
and farther away,
while the scene behind is expanded, redshifted, and closer.



\coordfig

\subsection{Affine distance in a Schwarzschild black hole}
\label{affinebh}

This section provides some techical details of ray-tracing
in the Schwarzschild
\cite{Schwarzschild,Schwarzschild:translation}
geometry,
used to construct the visualizations illustrated in \S\ref{sceneinside}.
Reading this section is not necessary to appreciate the visualizations.

A Schwarzschild black hole is the simplest kind of black hole,
a static, spherically symmetric black hole,
characterized entirely by its mass $M$.
The Schwarzschild metric in polar coordinates
$x^\mu \equiv \{ t , r , \theta , \phi \}$,
Figure~\ref{coord},
is,
in geometric units ($c = G = 1$),
\begin{equation}
  \dd s^2
  =
  - \, B \, \dd t^2
  + B^{-1} \dd r^2
  + r^2 ( \dd \theta^2 + \sin^2\!\theta \, \dd \phi^2 )
  \ ,
\end{equation}
where
\begin{equation}
  B \equiv 1 - 2 M / r
  \ .
\end{equation}
The vanishing of $B$ defines the horizon,
the Schwarzschild radius $r_s = 2 M$.
In the Schwarzschild metric,
the coordinate 4-velocity
$k^\mu \equiv \dd x^\mu / \dd \lambda$
along a null geodesic satisfies three conservation laws,
associated with
energy,
mass,
and
angular momentum $\bJ$ per unit energy,
\begin{equation}
\label{kschw}
  k^t = {1 / B}
  \ , \quad
  k^r = \pm \sqrt{1 - B J^2 / r^2}
  \ , \quad
  \bk^\perp = {\bJ / r^2}
  \ .
\end{equation}
Here
the photon energy $k^t$ has
been normalized to one as perceived by observers at rest at infinity.
Integrating $\dd r / \dd \lambda = k^r$ yields
the affine distance $\lambda$
between emitter and observer,
normalized
to observers at rest at infinity:
\begin{equation}
\label{lambdaschw}
  \lambda
  =
  \int_{r_\emit}^{r_\obs}
  {\dd r \over \sqrt{1 - B J^2 / r^2}}
  \ .
\end{equation}
The observed affine distance is then
\begin{equation}
\label{lambdaobs}
  \lambda_\obs
  =
  E_\obs \lambda
\end{equation}
where
$E_\obs$
is the photon energy perceived by the observer
relative to the photon energy perceived by observers at rest at infinity.
If the observer has specific energy $E$ and angular momentum $\bL$,
then their coordinate 4-velocity $u^\mu \equiv \dd x^\mu / \dd \tau$ is
\begin{equation}
  u^t = {E \over B}
  \ , \quad
  u^r = \pm \sqrt{E^2 - B ( 1 + L^2 / r^2 )}
  \ , \quad
  \bu^\perp = {\bL / r^2}
  \ .
\end{equation}
The observed photon energy
$E_\obs \equiv - u_\mu k^\mu$
is then
\begin{equation}
\label{Eobs}
  E_\obs
  =
  {E \mp \sqrt{( 1 - B J^2 / r^2 ) \left[ E^2 - B ( 1 + L^2 / r^2 ) \right]}
  \over B}
  -
  {\bL \cdot \bJ \over r^2}
  \ .
\end{equation}

The view from inside a black hole is most symmetrical
in the frame of an observer radially free-falling
on the zero-energy geodesic, $E = 0$, $L = 0$,
at the border between ingoing ($E > 0$)
and outgoing ($E < 0$).
An observer on the zero-energy radial geodesic sees
a photon of angular momentum $J$
subtend an angle $\chi$ away from the vertical axis
(see Figure~\ref{coord}) given by
\begin{equation}
\label{J0}
  | \tan \chi |
  =
  \sqrt{- B}
  J / r
  \ .
\end{equation}
The zero-energy radial observer sees
the observed photon energy $E_\obs$, equation~(\ref{Eobs}),
relative to its energy at rest at infinity
to be
\begin{equation}
\label{Eobs0}
  E_\obs
  =
  \sqrt{ {1 - B J^2 / r^2 \over - B} }
  =
  {| \sec\chi | \over \sqrt{- B}}
  \ .
\end{equation}

\section{Example: inside a Schwarzschild black hole}
\label{sceneinside}

Black holes are extreme examples of curved spacetime,
perfect for exploring with computer visualization
\cite{Mueller:PhD,Krasnikov:2008nj, Muller:2008zzk}.

\sceneoneframesfig

\subsection{Visualization}

Figure~\ref{scene1frames}
shows six frames from a visualization
\cite{Hamilton:website}
of the scene
seen by an observer falling into a Schwarzschild black hole.
The observer starts from outside the horizon,
falls on a free-fall trajectory,
passes through the horizon,
and terminates at the central singularity.

Figure~\ref{scenestereo}
shows frames
three to six of
Figure~\ref{scene1frames}
in stereo.
In these stereo views,
the distance from any point on the scene to the observer
has been set equal to the observed affine distance $\lambda_\obs$.
An animated version of the stereo visualization is at
\cite{Hamilton:website}.

The field of view
in Figures~\ref{scene1frames} and \ref{scenestereo}
is $105^\circ$ across the diagonal,
with a $4 \times 3$ aspect
($84^\circ \times 63^\circ$)
in the normal scene,
and a square aspect
($74.2^\circ \times 74.2^\circ$)
in the stereo scene.
In the stereo scene,
the eyes are $0.04 M$ geometric units ($0.02$ Schwarzschild radii) apart.
The background is Axel Mellinger's Milky Way
\cite{Mellinger},
reproduced here with permission.
The view direction is such as to put the radial axis
between the observer and the black hole's center
$6.32^\circ$ away from the center of the frame,
in a direction rotated $6.4^\circ$ clockwise from downward on the frame
(corresponding to a position of approximately 6.31 on a clockface).
Despite the appearance of later images of Figure~\ref{scene1frames}
that give the impression that the observer is looking sideways,
in fact the observer is always looking in the same direction,
approximately towards the black hole's center.
The deceptive appearance is caused by special relativistic aberration.

The observer is falling on a geodesic with specific energy and angular momentum
$E = 1$ and $L = 3.92 M$
geometric units
($c = G = 1$).
The angular momentum is slightly less than the angular momentum, $L = 4 M$,
that would put the observer on the $E = 1$ unstable circular orbit.
In the complete animated visualization,
the observer's orbit starts at $2060 M$ geometric units from the black hole,
in the plane of the Galaxy,
with the black hole positioned directly towards the Galactic Center.
The orbit is inclined at $5.7^\circ$ to the Galactic plane.
The grid on the black hole has its north pole tilted by $16^\circ$
relative to Galactic North Pole,
towards Galactic longitude $l = 322.1^\circ$.
The Figure shows the view from radii
$6.011 M$ (just outside the innermost stable circular orbit at $r = 6 M$),
$3.000 M$ (at the photon sphere at $r = 3 M$),
$1.997 M$ (just inside the horizon at $r = 2 M$),
$1.613 M$,
$0.786 M$,
and
$0.045 M$
(near the singularity at $r = 0$)
geometric units.

\scenestereofig

The apparent behavior of the horizon,
which appears to split into two
when an observer falls through it,
is quite unexpected.
A Penrose diagram,
Figure~\ref{bhholo_schw},
helps to navigate and make sense of the journey.
This diagram shows only two of our universe's four dimensions,
one time dimension (vertically) and one radial space dimension (horizontally).
A Penrose diagram is drawn so that
light always moves at $45^{\circ}$ from vertical.
An observer necessarily moves at less than the speed of light,
so follows an upward worldline
that is always less than $45^\circ$ from vertical.

\bhholoschwfig

As its Penrose diagram,
Figure~\ref{bhholo_schw}, shows,
the Schwarzschild geometry contains
both a true horizon,
and something that we call the illusory horizon.
In the analytically extended Schwarzschild geometry,
the illusory horizon is an actual horizon,
the horizon of a white hole and a parallel universe,
sometimes also called an antihorizon,
or past horizon\footnote{
We choose not to use the terms past and future horizon
because a future horizon becomes a past horizon
after you have passed through it.
}.
In a real black hole formed from the collapse of the core of a star,
there is no white hole horizon.
Rather, the horizon
is replaced by an exponentially redshifting
region that contains the collapsing star and its interior
(and beyond).
As time goes by,
the light escaping from the collapsing star becomes exponentially redshifted,
giving the illusory horizon its black appearance.
In this paper,
we suppose that the black hole collapsed long enough ago
that it has become effectively indistinguishable
from a Schwarzschild black hole,
and we refer to the exponentially redshifting surface
as the illusory horizon.
See \cite{Hamilton:collapse}
for animations that depict the idealized collapse of a spherical star
to a black hole, and spacetime and Penrose diagrams thereof.

When an observer outside the horizon of a real black hole
observes the horizon,
they are actually observing the illusory horizon,
the surface of the collapsed star,
exponentially dimmed to blackness.

The Penrose diagram,
Figure~\ref{bhholo_schw}, shows
that when the observer subsequently falls through the horizon,
they do not fall through the horizon they see, the illusory horizon.
Rather, they fall through the true horizon, which is not visible
to them until after they have passed through it.
Once inside the horizon,
the infaller sees both true and illusory horizons.
The illusory horizon
appears generally ahead of the infaller,
in the direction towards the black hole,
just as it did when the infaller was outside the horizon.
The true horizon
appears generally behind the infaller,
in the direction away from the black hole.

Inside the horizon,
all geodesics, including those of light rays,
necessarily fall to smaller radii.
It is consistent to conceive that space is falling superluminally
(faster than light) inside the horizon
\cite{Hamilton:2004au}.
Geodesics inside the horizon can be classified as
``ingoing'' or ``outgoing''
depending on whether they are moving backward or forward
in Schwarzschild coordinate time $t$,
that is,
on whether they have positive or negative energy\footnote{
Inside the horizon of a Schwarzschild black hole,
positive energy (ingoing) geodesics go
backwards in conventional Schwarzschild coordinate time $t$.
The sign comes from imposing
$t \rightarrow \infty$ at the true horizon,
and
$t \rightarrow - \infty$ at the illusory horizon.
}.
In the Penrose diagram,
Figure~\ref{bhholo_schw},
radial light rays moving to the left
inside the horizon are ingoing
(positive energy, going with the inflow of space),
while radial light rays moving to the right are outgoing
(negative energy, going against the inflow).
Outside the horizon all geodesics move forward
in Schwarzschild coordinate time $t$,
and can be classified as infalling or outfalling
depending on whether they are moving to smaller or larger radii.
For brevity, we use the terms ingoing and outgoing to refer also
to infalling and outfalling geodesics outside the horizon.

From outside the horizon of the black hole,
first two frames of Figure~\ref{scene1frames},
the observer sees only the illusory horizon.
The illusory horizon in
Figure~\ref{scene1frames}
is painted with a dark red grid,
as befits the fact that objects at this horizon
appear (disappear?) infinitely redshifted.
The illusory horizon appears to recede into the infinite distance,
as photons emanating from the horizon are able to orbit
the black hole an infinite number of times near the photon sphere
before peeling off into the observer's eye.

In frame two of Figure~\ref{scene1frames},
the observer is at the radius $r = 3 M$ of the photon sphere,
where light orbits in circles around the black hole.
The circular orbit is unstable,
and an observer at this radius
does not see light concentrate at the photon sphere.

Inside the horizon,
frames three to six of Figure~\ref{scene1frames},
the observer sees both illusory and true horizons.
The illusory horizon remains infinitely redshifted,
even though the observer is inside the horizon.
By contrast, an object falling through the true horizon
appears redshifted or blueshifted by a finite (not infinite) amount.
The true horizon in
Figure~\ref{scene1frames}
is painted with a grid in a blackbody color
that the observer would see an object at the horizon
to have if its emitting temperature were $4096 \unit{K}$
(peach white)
and it were free-falling radially from zero velocity at infinity
($E = 1$, $L = 0$).

\sceneonefig

\schwonestereofig


\scenefig

\bhholoschwzerofig

\schwzerostereofig

In frame three of Figure~\ref{scene1frames},
the observer is just inside the horizon.
At the instant that the infaller passes through the true horizon,
the true horizon appears not as a surface,
but rather as a line (which appears in projection to the observer as a point)
that extends from the position of the observer down to the illusory horizon.

It is sometimes asserted that an observer near the horizon
sees the outside universe concentrated into
a tiny, highly blueshifted, circular patch
of sky directly above them.
This would be true if the observer were at rest
in Schwarzschild coordinates,
but this is a highly unnatural situation,
requiring the observer to accelerate enormously just to remain at rest.
At and inside the horizon,
it is impossible for an observer to remain at rest,
since space is falling at or faster than the speed of light
\cite{Hamilton:2004au}.
The more realistic situation is that the observer falls,
as in the case illustrated by Figures~\ref{scene1frames}
and \ref{scenestereo}.

In frames three to six of Figure~\ref{scene1frames},
the observer is inside the horizon.
Together, the true and illusory horizons
form apparently 2D surfaces that encompass the infaller,
the ``Schwarzschild bubble.''
The illusory and true horizons join at a circle,
where the photon angular momentum $J$ is infinite.
More distant parts of the illusory horizon are visible
through the true horizon.
Light from the distant illusory horizon was emitted upward,
but turned around and fell back through the true horizon.

As the observer falls inward,
the Schwarzschild bubble expands horizontally.
In the limit as the observer approaches the singularity,
the edge of the Schwarzschild bubble extends to an angle of $\pi$
around the black hole,
so that the bubble finally encompasses
the full surface of each of the true and illusory horizons.

As the observer approaches the singularity,
the horizons appear to flatten out,
and appear to approach each other ever more closely,
as illustrated in
the final frame of Figure~\ref{scene1frames}.
The appearance is caused by the growing tidal field
near the singularity,
which aberrates the view so as to shift the apparent positions
of objects away from vertical towards horizontal.

\subsection{Scenes are constructed by Lorentz transforming out of radially falling frames}

Scenes seen by observers at the same position but with different motions
are related by a Lorentz transformation,
in the manner illustrated by Figure~\ref{beam}.
It is advantageous
to construct a scene in a frame where symmetries are manifest,
and then to Lorentz transform
the scene to the frame of the moving observer.
For example,
the visualizations
in Figures~\ref{scene1frames} and \ref{scenestereo}
were constructed by Lorentz transforming scenes
calculated from the perspective of a radially falling observer,
as described in the remainder of this subsection.

The spacetime of a spherical black hole has rotational symmetry
about the radial axis joing the center of the black hole and the observer.
Ray-tracing is simplified and accelerated in a frame where the
observer's motion is along the radial axis,
since the rotational symmetry about the axis
remains manifest in such a frame.

The rotational symmetry holds for any radial motion of the observer.
A natural choice of radial frame is that of an observer
who free-falls radially from zero velocity at infinity,
with specific energy $E = 1$ and angular momentum $L = 0$.
The radial free-fall frame exists at all radii,
both outside and inside the horizon.
Another possible choice of radial frame is the rest frame,
where the observer is at rest in Schwarzschild coordinates.
The rest frame fails at and inside the horizon,
since space is falling superluminally inside the horizon,
so no physical object can remain at rest.
A third choice of radial frame is the zero energy frame,
$E = L = 0$, which exists only inside the horizon.

Figure~\ref{scene1}
illustrates cross-sections through the effective 3-dimensional scene
perceived by an observer
who free-falls radially from zero velocity at infinity, with
$E = 1$, $L = 0$.
The Figure shows six successive views of
the location of the true and illusory horizons
of the Schwarzschild black hole.
The radial axis joining the black hole's center and the observer
lies along the vertical mid-line of the Figure.
The geometry is rotationally symmetric about the radial axis.

The third frame from top in Figure~\ref{scene1}
shows additional detail for the observer at a radius of $1 M$ geometric unit.
The Figure shows
the perceived location
of emitting surfaces at radii of
$3.5 M$, $3 M$, $2.5 M$, $2 M$, and $1.5 M$ geometric units.
Inside the horizon,
an observer can see points only at radii larger than their own:
points at smaller radii are in the future direction, and therefore invisible.
The observer sees
the surface at $1.5 M$ geometric units
to surround them.
Ingoing photons (blue) appear generally above the observer
(away from the black hole),
while outgoing photons (red) appear generally ahead of the observer
(towards the black hole).

Figure~\ref{schw1stereo}
shows a stereo visualization of the view depicted
in the third frame from top in Figure~\ref{scene1}.
The observer is
at a radius of $1 M$ geometric unit, inside the horizon.
The view shares some of the parameters of the scenes in
Figures~\ref{scene1frames} and \ref{scenestereo}:
the field of view is $105^\circ$ across the diagonal,
the orientation of the grid on the horizon is the same,
and the black hole is positioned directly in front of the Galactic Center,
which was the starting point of the trajectory in
Figures~\ref{scene1frames} and \ref{scenestereo}.
However, the view direction differs from
Figures~\ref{scene1frames} and \ref{scenestereo}:
the observer is looking in a direction $30^\circ$ below horizontal,
rotated left-handedly (scene shifts to the right)
by $54^\circ$ from Galactic North Pole
about the radial (vertical) direction.

The view from inside a black hole is most symmetrical
in the frame of an observer radially free-falling on the
zero-energy geodesic, $E = L = 0$,
at the border between ingoing ($E > 0$) and outgoing ($E < 0$).
Figure~\ref{scene}
shows cross-sections through the effective 3-dimensional scene
perceived by an observer on the radial zero-energy geodesic.
This is essentially the same set of views as Figure~\ref{scene1},
but from the perspective of the zero-energy geodesic.
The scenes in Figures~\ref{scene1} and \ref{scene}
are related by a radial Lorentz boost.
The zero-energy frame
reveals the symmetry between ingoing and outgoing geodesics,
evident in Figure~\ref{scene} as a reflection symmetry,
about a horizontal plane passing through the observer,
between the perceived locations of
ingoing (blue) and outgoing (red) surfaces.

Figure~\ref{bhholo_schw0}
shows the Penrose diagram of the black hole
with the observer falling on the zero-energy radial geodesic.
Figure~\ref{schw0stereo}
shows the scene seen by such an observer at a radius of
$1 M$~geometric unit,
the same as illustrated in
the second panel from top of Figure~\ref{scene}.
The scene in Figure~\ref{schw0stereo} is the same as in
Figure~\ref{schw1stereo},
Lorentz-boosted radially upward.
The observer in
Figure~\ref{schw0stereo}
is looking in the horizontal direction.
The true and illusory horizons show reflection symmetry
about the midplane of the view.
The symmetry does not extend to the coloration of the horizons,
since
objects are falling towards the observer through the true horizon above,
and away from the observer through the illusory horizon below.

\section{Perceptual distances}
\label{perceptual}

The affine distance may be the natural measure of distance
along a null geodesic in general relativity,
but
this distance is not directly measurable to an observer
using only local measurements.
In this section we therefore consider various perceptual distances
that are directly measurable by the observer.
We use the example of Section~\ref{sceneinside}
to compare perceptual distances to the affine distance,
Figure~\ref{dist}.

\subsection{Size distance}
\label{size}

The perceived angular size of an object
of known proper size provides one measure of distance,
known to astronomers as the angular diameter distance.
%
The size distance
requires prior knowledge of the object's actual size.

\magfig

%
%
%

\distfig

The size distance
$L_{\rm size}$
may be calculated by following a
narrow bundle of light rays
along the past lightcone of the observer
from the observer to the emitter,
as illustrated in Figure~\ref{mag}.
The bundle is characterized by
its cross-section in the 2-dimensional spatial plane
that lies in the 3-dimensional past lightcone of the observer,
and is orthogonal to the wavevector $k^\mu$.
The proper width of the bundle at any point along it is
$\delta l = \sqrt{\delta l^\nu \delta l_\nu}$,
where
$\delta l^\nu$
is an interval across the 2-dimensional cross-section.
The bundle starts at a point at the observer,
spreading out with angle $\delta \alpha$
\begin{equation}
\label{dalpha}
  \delta \alpha
  =
  \left. {\DD \delta l \over \DD \lambda} \right|_\obs
  \ ,
\end{equation}
where $\DD / \DD \lambda$
denotes covariant differentiation
with respect to affine distance $\lambda$.
As remarked at the end of the introduction to \S\ref{affine},
the affine distance here is understood to be
normalized to the rest frame of the observer,
so that it measures proper distances
in the local frame of the observer.
The size distance
$L_{\rm size}$
is the ratio of the width
$\delta l_\emit$
of the light bundle when it arrives at the emitter
to the angle
$\delta \alpha$
subtended by the bundle at the observer:
\begin{equation}
  L_{\rm size}
  =
  {\delta l_\emit \over \delta \alpha}
  \ .
\end{equation}
The size distance can be calculated by integrating
the equation of geodesic deviation along the null ray
from observer to emitter,
\begin{equation}
\label{geodesicdeviation}
  {\DD^2 \delta l^\nu \over \DD \lambda^2}
  =
  {R_{\kappa \lambda \mu}}^\nu k^\kappa k^\mu \delta l^\lambda
  \ ,
\end{equation}
subject to the initial conditions that
the spatial separation $\delta l^\nu$
lies in the light cone,
is orthogonal to the wavevector,
$k_\nu \delta l^\nu = 0$,
and is initially zero but with non-zero infinitesimal derivative
$\DD \delta l^\nu / \DD \lambda$,
equation~(\ref{dalpha}).
The equation of geodesic deviation~(\ref{geodesicdeviation}),
which depends on the Riemann curvature tensor
$R_{\kappa \lambda \mu \nu}$,
is a linear equation for the evolution of deviations $\delta l^\nu$
in the two-dimensional spatial plane transverse to the wavevector $k^\mu$.
The transformation matrix
$R_{\kappa \lambda \mu \nu} k^\kappa k^\mu$
is symmetric in $\lambda \nu$,
and the solution therefore has two orthogonal eigenvectors.
There are two corresponding size distances
in orthogonal directions.
Rather than integrate equation~(\ref{geodesicdeviation}) directly,
we prefer to evaluate $\delta l^\nu$ by ray-tracing along the null ray,
\begin{equation}
\label{dlnu}
  \delta l^\nu
  =
  \delta
  \int \dd x^\nu
  =
  \delta
  \int k^\nu \dd \lambda
  \ ,
\end{equation}
subject to the constraint
that the observed affine distance from observer to emitter is constant,
$\delta \left( E_\obs \int \dd \lambda \right) = 0$:
\begin{eqnarray}
\label{dlnuc}
  \delta l^\nu
  &=&
  \int \delta \left( {k^\nu \over E_\obs} \right) E_\obs \, \dd \lambda
\nonumber
\\
  &=&
  \int \delta k^\nu \, \dd \lambda
  -
  \delta \ln E_\obs \int k^\nu \, \dd \lambda
  \ .
\end{eqnarray}
The second term on the right hand side of equation~(\ref{dlnuc}),
involving the variation
$\delta \ln E_\obs$
of the observed photon energy in different directions
along the ray bundle,
appears because
in the Schwarzschild geometry
it is convenient to do the ray-tracing
with the wavevector $k^\nu \equiv \dd x^\nu / \dd \lambda$
and affine distance $\lambda$
normalized to observers at rest at infinity,
equations~(\ref{kschw}) and (\ref{lambdaschw})
(equation~(\ref{dlnu}) is valid regardless of normalization).
In this case the observed affine distance
is related to the rest-at-infinity affine distance
by equation~(\ref{lambdaobs}),
with observed energy $E_\obs$ from equation~(\ref{Eobs}).


The upper panel of
Figure~\ref{dist}
shows the reciprocal of the polar and azimuthal size distances
to the true and illusory horizons of the Schwarzschild black hole
as seen by the observer whose view is illustrated
in Figure~\ref{schw0stereo}
and in the second panel from top of Figure~\ref{scene}.
The two size distances differ,
but the average of the inverse size distances
provides a good estimate of the affine distance,
at least as long as the two size distances
are not too discrepant.

\subsection{Binocular distance}
\label{binocular}

The size distance has the drawback that the observer must
know in advance the actual size of the object being observed.
The size distance may be imprecise, or even completely wrong,
if the actual size fails to conform to the observer's expectation.

A measure of distance used successfully by
earthly animals
is the binocular distance,
also known to astronomers as parallax distance.
The binocular distance is the radius of curvature of
the 2-dimensional wavefront of light rays emitted by the emitter,
as measured by the observer in their own locally inertial rest frame,
as illustrated in Figure~\ref{bin}.
Binocular distance does not require prior knowledge of an object's actual size.
A different definition of binocular distance has been proposed by
\cite{Borchers:PhD},
but that definition does not yield a consistent
measure of distance in flat (Minkowski) space
(see \ref{borchers}).

\binfig

Question for the reader:
if you look at this page through a magnifying glass,
does your binocular vision perceive the text to be
closer or farther away?
Try it.
You can anticipate the answer by looking at Figure~\ref{bin}.

Consider two eyes a small distance $\delta l^\nu$ apart.
The covariant difference
$\delta k^\mu$
in the photon wavevectors
perceived by the two eyes is
\begin{equation}
\label{deltakmu}
  \delta k^\mu
  =
  \delta l^\nu {k^\mu}_{;\nu}
  \ ,
\end{equation}
where the colon $;$ represents covariant differentiation.
Figure~\ref{bin} illustrates the two wavevectors
$k^\mu$ and $k^\mu + \delta k^\mu$
seen by the two eyes.
Because the wavevector $k^\mu$, equation~(\ref{kmu}),
is normalized to unit energy,
the difference
$\delta k = \sqrt{\delta k^\mu \delta k_\mu}$
of the photon wavevectors
is just equal to the angle between the two wavevectors.
The resulting binocular distance $L_{\rm bin}$ is
\begin{equation}
  L_{\rm bin}
  =
  {\delta l \over \delta k}
  \ .
\end{equation}
The eye separation $\delta l^\nu$
is confined to lie on the 3-dimensional future lightcone of the emitter,
and the geodesic equation
$D k^\mu / D \lambda = k^\nu {k^\mu}_{;\nu} = 0$
implies that $\delta k^\mu$ vanishes
for $\delta l^\nu$ parallel to the wavevector $k^\nu$.
Thus the covariant difference
$\delta k^\mu$
depends only on the components of the eye separation
in the 2-dimensional spatial subsurface of the emitter lightcone
that is orthogonal to the wavevector,
and without loss of generality the eyes can be taken to lie
in this 2-dimensional subsurface.
More explicitly,
the covariant difference
in photon wavevectors between the two eyes is
\begin{equation}
\label{deltakmuGamma}
  \delta k^\mu
  =
  \delta l^\nu
  \left(
  {\partial k^\mu \over \partial x^\nu}
  +
  \Gamma^\mu_{\kappa\nu} k^\kappa
  \right)
  \ ,
\end{equation}
in which
the first term,
$\delta l^\nu \partial k^\mu / \partial x^\nu$,
represents the change in the photon wavevector
$k^\mu$
between the two eyes,
while the second term,
$\delta l^\nu \Gamma^\mu_{\kappa\nu} k^\kappa$,
depending on the connection coefficients
$\Gamma^\mu_{\kappa\nu}$,
represents the effect of parallel-transporting
the photon wavevector from one eye to the other.
The second term corrects
for the difference in spacetime frames of the two eyes
that results from the arbitrariness of coordinates in general relativity.
It is only the covariant difference in photon wavevectors
that is physically measurable.
Evaluating the first term
on the right hand side of equation~(\ref{deltakmuGamma})
involves ray-tracing
from emitter to observer,
while the second term on the right depends only on quantities
local to the observer.

In geometric optics
\cite[\S22.5]{MTW:1973},
the future lightcone of the emitter is a surface of constant phase
$\psi$,
the wavefront.
The photon wavevector $k^\mu$
is given by the gradient of the phase,
$k_\mu = \psi_{;\mu} = \partial \psi / \partial x^\mu$.
The null condition
$k^\mu k_\mu = 0$
implies that the photon wavevector $k^\mu$ is orthogonal
to the gradient $k_\mu$ of the phase,
and therefore lies in the lightcone, the surface of constant phase.
The covariant derivative
of the photon wavevector
$k_{\mu;\nu}$
is the curvature of the phase,
a symmetric matrix:
\begin{equation}
  k_{\mu;\nu}
  =
  \psi_{;\mu\nu}
  \ .
\end{equation}
In the locally inertial frame of the observer,
and on the lightcone,
the non-vanishing components of the curvature matrix
$\psi_{;\mu\nu}$
constitute a $2 \times 2$ symmetric matrix
in the spatial directions orthogonal to the wavevector.
The eigenvalues of the symmetric matrix
$\psi_{;\mu\nu}$
are the reciprocal radii of curvature
$1/L_\parallel$ and $1/L_\perp$
in two perpendicular directions.
In a frame aligned with the eigenvectors,
the covariant difference $\delta k^\mu$,
equation~(\ref{deltakmu}),
in the photon wavevectors seen by two eyes becomes
\begin{equation}
\label{dkparperp}
  {1 \over E_\obs}
  \left(
  \begin{array}{c}
    \delta k^\parallel \\
    \delta k^\perp
  \end{array}
  \right)
  =
  \delta l
  \left(
  \begin{array}{cc}
    1 / L_\parallel & 0 \\
    0 & 1 / L_\perp
  \end{array}
  \right)
  \ .
\end{equation}
As with the size distance equation~(\ref{dlnuc}),
the factor of $E_\obs$ is introduced into equation~(\ref{dkparperp})
because in the Schwarzschild geometry it is convenient
to do the ray-tracing demanded by equation~(\ref{deltakmuGamma})
with the wavevector $k^\mu$ normalized to rest-at-infinity observers,
equation~(\ref{kschw}).
The wavevector must then be rescaled to the actual observer
with the observed energy $E_\obs$ from equation~(\ref{Eobs}).
In the Schwarzschild geometry,
the two radii of curvature $L_\parallel$ and $L_\perp$ of the wavefront
correspond to those with eyes separated in the polar and azimuthal directions,
as indicated in the diagram to the right side of Figure~\ref{dist}.

In flat spacetime (Minkowski space),
the wavefront from a point emitter is spatially spherical,
and the radius of curvature $L$ is the same,
and equal to the affine distance,
regardless of how the eyes are oriented about the line of sight.

In curved spacetime,
the curvature matrix has two different eigenvalues,
so the wavefront is ellipsoidal,
yielding two different measures of binocular distance
$L_\parallel$ and $L_\perp$.
The conflicting visual cues would disorient
an animal with binocular vision,
just as conflicting visual and motion cues disorient people
inside immersive environments such as airplanes.


The lower panel of
Figure~\ref{dist}
shows the reciprocal of the polar and azimuthal binocular distances
to the true and illusory horizons of a Schwarzschild black hole,
as seen by the observer whose view is shown in
in Figure~\ref{schw0stereo}
and in the second-from-top panel of
Figure~\ref{scene}.
The observer is inside the horizon at a radius of $1 M$ geometric unit,
and is radially infalling on the zero-energy geodesic.
On-axis ($0^\circ$ and $180^\circ$),
the binocular distances agree with the affine distance,
but off-axis the polar and azimuthal binocular distances disagree,
a symptom of the fact that
the wavefront is ellipsoidal rather than spherical.

The observer can also deduce an absolute distance
from the fractional difference in size distances
perceived by two eyes separated along the line of sight.
However, this distance is simply the binocular distance,
and so does not provide an additional measure of distance.


\subsection{Trinocular distance}
\label{trinocular}

One practical solution to the problem of depth perception
in curved space would be to evolve three eyes in a triangular
(non-collinear) pattern,
such as illustrated in Figure~\ref{ape}.
Three eyes would allow an animal to process the ellipsoidal wavefront
into a best distance,
one that would improve their chance of survival.


The lower panel of
Figure~\ref{dist}
shows that the mean of the reciprocal polar and azimuthal binocular distances
agrees well with the affine distance even well off-axis
where the wavefront is quite non-spherical.
This suggests that
a good strategy for the brains of three-eyed animals
would be to infer a trinocular distance
$L_{\rm tri}$
from the mean of the reciprocal binocular distances,
that is,
from the trace of the wavefront curvature matrix:
\begin{equation}
\label{Ltri}
  {1 / L_{\rm tri}}
  =
  {\textstyle \frac{1}{2}}
  \left(
  {1 / L_\parallel}
  +
  {1 / L_\perp}
  \right)
  \ .
\end{equation}

\apefig

The agreement of the
trinocular distance $L_{\rm tri}$, equation~(\ref{Ltri}),
with the affine distance
can be regarded as a consequence of the Raychaudhuri equation
\cite{Chandrasekhar:1983,Kar:2006ms},
which shows that in empty
space (zero energy-momentum tensor),
the expansion of a bundle of light rays depends only quadratically
on the shear.
Thus the expansion is unaffected to linear order by the shear.
In other words,
the expansion is more or less what it would be in the absence of shear,
provided that the ellipticity of the ray bundle
is less than unity.

The trinocular distance fails as an estimate of affine distance
when the ellipticity of the ray bundle exceeds unity.
The wise animal would learn not
to trust the trinocular distance
if the two binocular distances are too discrepant.

As a brief reminder of where the Raychaudhuri and related equations
come from,
a bundle of light rays may be characterized by the four Sachs scalars,
the expansion, the rotation (or vorticity), and complex shear
\cite[\S9]{Chandrasekhar:1983}, \cite{Kar:2006ms}.
The Sachs scalars are equivalent to the
$2 \times 2$ matrix
of tetrad-frame connection coefficents
orthogonal to the wavevector
evaluated in a sequence of locally inertial frames
parallel-transported along the null ray.
The evolution of the Sachs scalars is governed
by the usual equations relating the tetrad-frame connection coefficients
to the Riemann tensor.
In geometric optics, where the wavevector is the gradient of a scalar,
the rotation vanishes.
If the tidal stress (spin-$2$ component of the Weyl tensor)
along the bundle is non-vanishing,
then shear accumulates along the bundle,
causing the wavefront to become ellipsoidal.
The equation for the expansion,
the Raychaudhuri equation,
depends quadratically on the shear.

\section{Conclusion}
\label{summary}

A scene
in any arbitrary curved spacetime
may be rendered consistently in 3D
by placing objects
at their observed angular positions and
at their observed affine distances.
Such a scene may be displayed to an audience
in stereo in the usual fashion,
by projecting two views of the scene,
as if it were in flat space,
as seen by two cameras a small distance apart.

In flat (Minkowksi) spacetime,
affine distance coincides with proper distance
measured in the rest frame of the observer,
which in turn coincides with the binocular distance
perceived by a two-eyed observer,
even if the observer is moving through the scene at near the speed of light.

In curved spacetime, however,
the 3D stereo scene obtained by placing objects at
their affine distances
is {\em not\/} the same as the scene that an observer with
binocular vision would perceive if they were actually
present in the curved spacetime.

The fact that binocular vision fails in curved spacetime
is not a fundamental obstacle to 3D perception.
Rather, it is simply a limitation of beings
who have evolved in flat spacetime.
In flat spacetime,
wavefronts of light from an emitter
propagate outward spherically,
and binocular vision,
which measures the radius of curvature of the wavefront,
yields a unique estimate of distance
that coincides with affine distance.
But in curved spacetime,
tidal forces distort a wavefront from sphericity,
causing it to become locally ellipsoidal,
with two distinct curvature eigenvalues,
hence two conflicting measures of binocular distance.
Three eyes,
Figure~\ref{ape},
would allow an animal
to process the ellipsoidal wavefront into
an improved
estimate,
the trinocular distance.
It follows from the Raychaudhuri equation that
the average of the two inverse radii of curvature
would provide a trinocular distance that is a good estimate of the affine distance
provided that the ellipticity of the wavefront is not much greater than one.

\section*{Acknowledgments}

This work was supported in part by NSF award
AST-0708607.
We thank Wildrose Hamilton
(\url{http://www.wildrosehamilton.com})
for
Figure~\ref{ape}.

\appendix

\section{Rigid bodies in curved spacetimes}
\label{rigid}

It is impossible to
place a flat piece of paper on the surface of a sphere
without wrinkling the paper.
For essentially the same reason,
there is no such thing as a rigid body in curved spacetime.
The difficulty arises only if the size of the body
is appreciable compared to the length scale of curvature of the spacetime.
General relativity postulates the existence of locally inertial frames,
and if the body is small enough,
then it will fit in a locally inertial frame with negligible distortion.

In the scene shown in Figure~\ref{earthbinocular},
the Earth is placed artificially, as if it were a rigid body,
near a Schwarzchild black hole.
This Appendix describes the convention adopted here
for placing a large rigid body in Schwarzschild spacetime.

The Gullstrand-Painlev\'e
\cite{Gullstrand:1922,Painleve:1921,Hamilton:2004au}
version of the Schwarzschild line-element
for a spherical black hole of mass $M$ is
\begin{equation}
\label{GPlineelement}
  \dd s^2
  =
  - \,
  \dd t_\ff^2
  +
  ( \dd r - v \, \dd t_\ff )^2
  +
  r^2 ( \dd \theta^2 + \sin^2\!\theta \, \dd \phi^2 )
  \ ,
\end{equation}
where $t_\ff$ is the free-fall time,
the proper time attached to observers who free-fall
from zero velocity and angular momentum at infinity,
and $v$ is the Newtonian escape velocity
\begin{equation}
  v
  =
  -
  \sqrt{2 M \over r}
  \ .
\end{equation}
The Gullstrand-Painlev\'e line-element~(\ref{GPlineelement})
has the property that it is spatially flat at constant free-fall time
$\dd t_\ff = 0$.
In a neighborhood of any point,
Gullstrand-Painlev\'e coordinates
define a set of locally inertial coordinates
attached to radial free-fall observers.
Thus a consistent, albeit somewhat arbitrary, approach
is to treat Gullstrand-Painlev\'e coordinates as defining
a mapping between Schwarzschild and Minkowski spacetime;
that is, the coordinates $\{ t_\ff , r , \theta , \phi \}$
are treated as polar coordinates of Minkowski space.
For example,
in the simple case of a body whose center-of-mass
is at rest in the frame of a Gullstrand-Painlev\'e free-faller,
the body would be placed in the flat Gullstrand-Painlev\'e
spatial hypersurface as if the space were plain 3-dimensional Euclidean space.
In the case shown in Figure~\ref{earthbinocular},
the Earth is in orbit around the Schwarzschild black hole,
so its frame is boosted relative to Gullstrand-Painlev\'e free-fallers.

\section{Alternative definition of binocular distance}
\label{borchers}

In this paper, the binocular distance from an emitting source
has been taken to be the radius of curvature of the 2-dimensional wavefront
from the source,
as perceived by an observer in their own locally inertial frame.
In flat (Minkowski) space,
this yields a consistent measure of distance
regardless of the motion of the emitter or observer.

Chapter~3 of \cite{Borchers:PhD}
considers an alternative definition in which
a binocular scene is constructed from two images
viewed simultaneously in the rest frame of the observer.
By contrast,
the definition in the present paper
takes the two images to lie in the lightcone of the emitter,
in which case there is a slight time delay between the two images,
equal to the differential travel time of the light to the two eyes.
The definitions of \cite{Borchers:PhD} and the present paper
agree when the eyes are looking straight ahead, so that
there is no differential time delay in the light travel time to the eyes,
but differ when the eyes are not looking directly ahead.

As
\cite{Borchers:PhD}
points out, two simultaneous images
do not yield a consistent stereo view.
By contrast,
the present definition of binocular distance
as the radius of curvature of the wavefront
yields a consistent measure of distance,
and thus a consistent stereo view, in Minkowski space.
Successful binocular vision would
compensate for the differential light travel time between the two eyes.
Such compensation would be quite feasible,
since it depends only on the distance between the eyes
and the observed angle of the emitter relative to directly ahead.

Human vision is too slow to distinguish images separated by
a light travel time between the eyes,
which is less than a nanosecond.
More precise vision is needed
to achieve successful binocular vision in a relativistic setting.

Astronomers are familiar with the fact that precision
location of objects on the sky (astrometry)
requires faithful tracking of the wavefront.
For example,
in Very Long Baseline Interferometry
\cite{Burke:2009}
images seen by two or more telescopes
are combined with an appropriate time delay
so that all see the same wavefront.
The Keck telescope,
consisting of twin 10-meter optical/infra-red telescopes
on the summit of Mauna Kea in Hawai'i,
can be operated in interferometric mode
\cite{Ragland:2010}.
The Keck telescope thus realizes the requirements
for successful binocular vision in a relativistic environment.

\section*{References}

\bibliographystyle{unsrt}

\bibliography{bh}

\begin{thebibliography}{10}

\bibitem{Ruder:2008}
Hanns Ruder, Daniel Weiskopf, Hans-Peter Nollert, and Thomas M{\"u}ller.
\newblock How computers can help us in creating an intuitive access to
  relativity.
\newblock {\em New J.\ Phys.}, 10:125014, 2008.

\bibitem{10.1109/TVCG.2006.69}
D.~Weiskopf, M.~Borchers, T.~Ertl, M.~Falk, O.~Fechtig, R.~Frank, F.~Grave,
  A.~King, U.~Kraus, T.~M{\"u}ller, H.-P. Nollert, I.~Rica Mendez, H.~Ruder,
  T.~Schafhitzel, S.~Sch{\"a}r, C.~Zahn, and M.~Zatloukal.
\newblock Explanatory and illustrative visualization of special and general
  relativity.
\newblock {\em IEEE Transactions on Visualization and Computer Graphics},
  12(4):522--534, 2006.

\bibitem{Hamilton:bhfs}
Andrew J.~S. Hamilton.
\newblock Black hole flight simulator.
\newblock \url{http://jila.colorado.edu/~ajsh/insidebh/bhfs.html}, 2010.

\bibitem{Mellinger}
Axel Mellinger.
\newblock All-sky {M}ilky {W}ay panorama.
\newblock \url{http://home.arcor-online.de/axel.mellinger/}, 2000.

\bibitem{Hamilton:website}
Andrew J.~S. Hamilton.
\newblock Journey into a {S}chwarzschild black hole.
\newblock \url{http://jila.colorado.edu/~ajsh/insidebh/schw.html}, 2009.

\bibitem{MTW:1973}
Charles~W. Misner, Kip~S. Thorne, and John~A. Wheeler.
\newblock {\em Gravitation}.
\newblock W. H. Freeman and Co., 1973.

\bibitem{Schwarzschild}
Karl Schwarzschild.
\newblock {\"U}ber das {G}ravitationsfeld eines {M}assenpunktes nach der
  {E}insteinschen {T}heorie.
\newblock {\em Sitzungsberichte der {D}eutschen {A}kademie der {W}issencshaften
  zu {B}erlin, {K}lasse f{\"u}r {M}athematik, {P}hysik, und {T}echnik},
  1916:189--196, 1916.

\bibitem{Schwarzschild:translation}
Karl Schwarzschild.
\newblock On the gravitational field of a mass point according to {E}instein's
  theory.
\newblock {\em Gen.\ Rel.\ Grav.}, 35:951--959, 2003.
\newblock English translation of {\cite{Schwarzschild}}.

\bibitem{Mueller:PhD}
Thomas M{\"u}ller.
\newblock Visualisierung in der {R}elativit{\"a}tstheorie.
\newblock Dissertation, Univ. Tuebingen,
  \url{http://nbn-resolving.de/urn:nbn:de:bsz:21-opus-24564}, 2006.

\bibitem{Krasnikov:2008nj}
S.~Krasnikov.
\newblock Falling into the {S}chwarzschild black hole. {I}mportant details.
\newblock {\em Grav. Cosmol.}, 14:362, 2008.

\bibitem{Muller:2008zzk}
Thomas M{\"u}ller.
\newblock Falling into a {S}chwarzschild black hole.
\newblock {\em Gen. Rel. Grav.}, 40:2185--2199, 2008.

\bibitem{Ghez:2003qj}
A.~M. Ghez, S.~Salim, S.~D. Hornstein, A.~Tanner, J.~R. Lu, M.~Morris, E.~E.
  Becklin, and G.~Duchene.
\newblock Stellar orbits around the galactic center black hole.
\newblock {\em Astrophys. J.}, 620:744--757, 2005.

\bibitem{Eisenhauer:2005cv}
Frank Eisenhauer, R.~Genzel, T.~Alexander, R.~Abuter, T.~Paumard, T.~Ott,
  A.~Gilbert, S.~Gillessen, M.~Horrobin, S.~Trippe, H.~Bonnet, C.~Dumas,
  N.~Hubin, A.~Kaufer, M.~Kissler-Patig, G.~Monnet, S.~Stroebele, T.~Szeifert,
  A.~Eckart, R.~Schoedel, and S.~Zucker.
\newblock {SINFONI} in the {G}alactic center: young stars and {IR} flares in
  the central light month.
\newblock {\em Astrophys. J.}, 628:246--259, 2005.

\bibitem{Hamilton:collapse}
Andrew J.~S. Hamilton.
\newblock Collapse to a black hole.
\newblock \url{http://casa.colorado.edu/~ajsh/collapse.html}, 1998.

\bibitem{Hamilton:2004au}
Andrew J.~S. Hamilton and Jason~P. Lisle.
\newblock The river model of black holes.
\newblock {\em Am. J. Phys.}, 76:519--532, 2008.

\bibitem{Borchers:PhD}
Marc~P Borchers.
\newblock Interaktive und stereoskopische {V}isualisierung in der speziellen
  {R}elativit{\"a}tstheorie.
\newblock Dissertation, Univ. Tuebingen,
  \url{http://nbn-resolving.de/urn:nbn:de:bsz:21-opus-18918}, 2005.

\bibitem{Chandrasekhar:1983}
Subrahmanyan Chandrasekhar.
\newblock {\em The mathametical theory of black holes}.
\newblock Clarendon Press, 1983.

\bibitem{Kar:2006ms}
Sayan Kar and Soumitra SenGupta.
\newblock The {R}aychaudhuri equations: {A} brief review.
\newblock {\em Pramana}, 69:49, 2007.

\bibitem{Gullstrand:1922}
Allvar Gullstrand.
\newblock Allgemeine {L\"o}sung des statischen {E}ink{\"o}rperproblems in der
  {E}insteinschen {G}.
\newblock {\em Arkiv.\ Mat.\ Astron.\ Fys.}, 16(8):1--15, 1922.

\bibitem{Painleve:1921}
Paul Painlev{\'e}.
\newblock La m{\'e}canique classique et la th{\'e}orie de la relativit{\'e}.
\newblock {\em C.\ R.\ Acad.\ Sci.\ (Paris)}, 173:677--680, 1921.

\bibitem{Burke:2009}
Bernard~F. Burke and Francis Graham-Smith.
\newblock {\em An Introduction to Radio Astronomy, 3rd edition}.
\newblock Cambridge University Press, 2009.

\bibitem{Ragland:2010}
S.~Ragland, R.~Akeson, M.~Colavita, R.~Millan-Gabet, J.~Woillez, P.~Wizinowich,
  E.~Appleby, B.~Berkey, A.~Cooper, C.~Felizardo, J.~Herstein, M.~Hrynevych,
  D.~Medeiros, D.~Morrison, T.~Panteleeva, J.-U. Pott, B.~Smith, K.~Summers,
  K.~Tsubota, C.~Tyau, and E.~Wetherell.
\newblock Recent progress at the {K}eck interferometer.
\newblock {\em Proc. SPIE}, 7734:773402, 2010.

\end{thebibliography}

\end{document}